\documentclass[a4paper,11pt]{article}
\usepackage{jheppub}
\usepackage{amsmath,amssymb}
\usepackage[T1]{fontenc}
\usepackage{charter}
\usepackage[normalem]{ulem} 
\numberwithin{equation}{section}
\title{Hironaka decompositions in supersymmetric theories: an exploration}
\author{Abhijit Gadde,}
\author{Rajat Raj,}
\author{and Shivansh Tomar}
\affiliation{Department of Theoretical Physics,\\
Tata Institute of Fundamental Research, Homi Bhabha Road, Mumbai 400005, India}
\emailAdd{abhijit@theory.tifr.res.in}
\emailAdd{rajat.raj@tifr.res.in}
\emailAdd{shivansh.tomar@tifr.res.in}
\hypersetup{
pdftitle={Hironaka decompositions in supersymmetric theories: an exploration},
pdfauthor={Abhijit Gadde, Rajat Raj, Shivansh Tomar}
}
\abstract{
We ask when tree-level supersymmetric cohomology admits a Hironaka decomposition: a unique expansion in finitely many secondary operators with polynomial coefficients in algebraically independent bosonic primaries. Such a decomposition organizes the cohomology into towers with a common set of unrestricted dressings. For diagonal models with $N$ labels, $d\geq1$ bosonic species and $f$ fermionic species, we construct this decomposition and obtain $(N!)^{d-1}2^{fN}$ secondaries. We also prove that the scalar-branch ring of ABJM theory admits a Hironaka decomposition at every rank and positive integer Chern--Simons level, and construct explicit bases for the $SU(2)\times SU(2)$ Klebanov--Witten scalar ring. We also do the same for the Higgs branches of two class-S theories. However, not all supersymmetric cohomologies admit a Hironaka decomposition. In the Wess--Zumino model with $W=x^2y^2$, every positive-degree scalar annihilates the nonzero class $xy$, while $x^n$ is in cohomology for every $n\geq 1$. These two properties rule out a Hironaka decomposition over scalar primaries.
}

\begin{document}
\maketitle

\section{Introduction}
\label{sec:intro}

The tree-level cohomology of a supersymmetric field theory carries a ring structure\footnote{By tree-level cohomology we mean the cohomology of the classical interacting supercharge. In $\mathcal N=4$ super-Yang--Mills theory, its classes correspond to states that remain BPS at one loop~\cite{Grant:2008sk}, hence the terminology ``one-loop cohomology'' used in~\cite{Gadde:2025yoa}. Here we use the more standard  ``tree-level cohomology,'' as in~\cite{Chang:2025mqp}, to specify the order of the supercharge rather than the dilatation operator.}. Choose a supercharge $Q$ that squares to zero on gauge-invariant operators. Its classical action contains a single functional derivative and therefore obeys the graded Leibniz rule,
\begin{equation}
    Q(AB)=(QA)B+(-1)^{|A|}A(QB),
\end{equation}
where $|A|$ is the fermion parity of $A$. Products of $Q$-closed operators are closed, and the product of a closed operator with a $Q$-exact operator is exact. Multiplication thus descends to cohomology. We ask how this ring can be organized. 

Following recent work on finite-rank matrix invariants~\cite{deMelloKoch:2025ngs} and their Hilbert spaces~\cite{deMelloKoch:2025rkw}, we ask whether supersymmetric cohomology admits a Hironaka decomposition. In this organization, one chooses algebraically independent bosonic operators $p_i$, called primaries, and finitely many operators $s_\alpha$, called secondaries, such that every cohomology class has a unique expression
\begin{equation}
    \mathcal O=\sum_\alpha s_\alpha F_\alpha(p_1,\ldots,p_m),
\end{equation}
with polynomial coefficients $F_\alpha$. Each secondary therefore generates a tower under multiplication by the same primaries. In a bulk description, the secondaries might describe background states, including black-hole states, and the primaries might generate graviton excitations around them. A relation between secondary invariants and nonperturbative sectors has been explored in zero-dimensional matrix integrals~\cite{deMelloKoch:2026secondary}. Establishing a corresponding interpretation for supersymmetric cohomology would require identifying both the backgrounds and the operators that generate their excitations.

In supersymmetric gauge theories, relations among operators have two sources. Finite-rank matrix identities relate trace expressions before taking cohomology. Supersymmetry supplies further relations by making combinations of closed operators \(Q\)-exact. The \(1/4\)-BPS sector and a particular \(1/8\)-BPS sector of \(\mathcal N=4\) super-Yang–Mills theory provide a simple setting in which both occur. The first contains two scalars and one fermion; the second contains three scalars and two fermions. All graded commutators vanish in cohomology, and both sectors admit descriptions by polynomials in bosonic and fermionic diagonal entries, invariant under simultaneous permutations of the \(N\) labels \cite{Kinney:2005ej}. The problem therefore reduces to constructing \(S_N\) invariants.

For commuting bosonic matrices, the  construction of~\cite{deMelloKoch:2025eqt} identifies the traces of powers of each matrix as primaries and gives an algorithm for constructing secondaries from trace relations; for two bosonic species, their number is $N!$. We extend this organization to diagonal models with $d\geq 1$ bosonic species and $f$ fermionic species. We prove that the number of secondaries is $(N!)^{d-1}2^{fN}$.

We give a general construction of the secondaries in Appendix~\ref{app:diagonal} and also construct explicit bases at low rank.  They are built from traces of products of different matrices, including fermionic matrices. We recover the same secondary count from the grand canonical partition function of the corresponding oscillator states.

The non-derivative cohomology of the Wess--Zumino model with $W=xyz$, including fermions, admits a Hironaka decomposition with one scalar primary and six secondaries. We also construct explicit Hironaka bases for the scalar cohomology of $SU(2)\times SU(2)$ Klebanov--Witten theory and for two class-S Higgs branches. We prove that the ABJM scalar-branch ring admits a Hironaka decomposition at every rank and positive integer Chern--Simons level. We also study a complex consisting of certain bosonic fields and a fermionic field, and show that its cohomology admits a Hironaka decomposition.

In the BMN truncation of $\mathcal N=4$ super-Yang--Mills theory, we construct a Hironaka decomposition for the bosonic diagonal image of the $SU(2)$ graviton ring. Whether this decomposition extends to the full graviton cohomology, including fermions, remains open. We also leave open whether operators outside the graviton sector admit a decomposition under multiplication by graviton operators. 

A Hironaka decomposition does not exist in every example of supersymmetric cohomology. In the polynomial cohomology of the Wess--Zumino model with superpotential $W=x^2y^2$, the nonzero class $xy$ is annihilated by every positive-degree scalar polynomial. No scalar primary can therefore act freely on the entire cohomology. At the same time, arbitrarily high powers of $x$ are in cohomology, so the cohomology cannot be spanned by finitely many secondaries without scalar dressings. Together, these facts rule out a Hironaka decomposition over scalar primaries. Table~\ref{tab:summary} summarizes our results and gives the number of primaries and secondaries wherever such a decomposition exists.

\begin{table}[!htbp]
\centering
\small
\renewcommand{\arraystretch}{1.15}
\begin{tabular}{@{}p{0.25\textwidth}p{0.43\textwidth}p{0.25\textwidth}@{}}
\hline
\raggedright Example &\raggedright  Operator space &\raggedright  Primaries; secondaries\tabularnewline\hline
\raggedright Diagonal models (Section~\ref{sec:commuting}) &\raggedright 
Permutation-invariant polynomials with $d$ bosonic and $f$ fermionic
species, including the $1/4$-BPS and $1/8$-BPS models &\raggedright 
$dN$; $(N!)^{d-1}2^{Nf}$\tabularnewline[5pt]
\raggedright Wess--Zumino, $W=xyz$ &\raggedright 
Polynomial cohomology, including fermions &\raggedright 
$1$; $6$\tabularnewline[5pt]
\raggedright Klebanov--Witten, $SU(2)\times SU(2)$ &\raggedright 
Scalar F-term ring &\raggedright 
$7$; $10$\tabularnewline[5pt]
\raggedright Class S, four-punctured $A_1$ sphere &\raggedright 
Higgs-branch scalar ring &\raggedright 
$10$; $168$\tabularnewline[5pt]
\raggedright Class S, closed genus-two surface &\raggedright 
Higgs-branch scalar ring &\raggedright 
$2$; $2$\tabularnewline[5pt]
\raggedright ABJM, rank $N$ and level $k$ &\raggedright 
Scalar-branch ring, for every $N\geq1$ and $k\geq1$ &\raggedright 
$4N$; $k^{3N}(N!)^3$\tabularnewline[5pt]
\raggedright Restricted bifundamental complex (Section~\ref{sec:restricted}) &\raggedright 
Cohomology of the specified $SU(2)\times SU(2)$ invariant complex &\raggedright 
$5$; $12$ (distributed as $3+6+3$ by fermion number)\tabularnewline[5pt]
\raggedright $SU(2)$ BMN graviton ring &\raggedright 
Bosonic diagonal image of graviton operators &\raggedright 
$4$; $21$\tabularnewline[5pt]\hline
\raggedright Wess--Zumino, $W=x^2y^2$ &\raggedright 
Polynomial cohomology: the nonzero class $xy$ is annihilated by every
positive-degree scalar polynomial &\raggedright 
No decomposition over scalar primaries\tabularnewline\hline
\end{tabular}
\caption{Summary of our main results. }
\label{tab:summary}
\end{table}

The cohomology of the $W=x^2y^2$ model does admit a well-studied generalization called a Stanley decomposition. It allows different seed classes to have different polynomial dressings while preserving the uniqueness of the expansion. In this example, $xy$ occurs with no dressings, alongside classes that admit infinite towers of dressing. Stanley-type decompositions may offer the flexibility required for organizing tree-level cohomology across supersymmetric gauge theories. Here we study Hironaka decompositions and leave this broader possibility for future work. 

The paper is organized as follows. In Section \ref{sec:hironaka}, we give a brief introduction to Hironaka decomposition. In Section \ref{sec:commuting}, we prove its existence for permutation-invariant diagonal models containing bosonic and fermionic variables. In Section \ref{sec:other}, we study several supersymmetric theories, constructing explicit Hironaka decompositions and presenting an example where such a decomposition fails to exist. In Section \ref{sec:graviton}, we construct a Hironaka decomposition for the bosonic diagonal image of the BMN graviton ring. Section \ref{sec:discussion} summarizes our results and discusses future directions. The appendices provide explicit sets of primaries and secondaries, together with supporting derivations. The codes to verify secondary relations are attached as ancillary files.  

\section{\texorpdfstring{Finite-$N$ relations and Hironaka decomposition}{Finite-N relations and Hironaka decomposition}}
\label{sec:hironaka}

Consider the half-BPS operators built from a single complex adjoint scalar $X$ in $U(N)$ $\mathcal N=4$ super-Yang--Mills theory. With no derivatives or other fields, these operators are products of traces of powers of $X$~\cite{Corley:2001zk,Berenstein:2004kk}. At finite $N$, however, not all traces are independent. The Cayley--Hamilton identity, which states that an $N\times N$ matrix satisfies its own characteristic equation, allows every trace of a power higher than $N$ to be written as a polynomial in
\begin{equation}
    p_k=\operatorname{Tr}X^k,\qquad k=1,\ldots,N.
\end{equation}
These traces are algebraically independent and generate all symmetric polynomials in the eigenvalues of $X$.

Every operator in this sector therefore has a unique polynomial expression in the $p_k$. This is the simplest Hironaka decomposition: the $p_k$ are the primaries, and the identity is the only secondary. Finite $N$ limits the number of independent trace generators but places no bound on how many times each primary can appear. For two commuting $U(2)$ scalars, mixed traces contain information not determined by the individual spectra.

Let $X$ and $Y$ be two $2\times 2$ commuting scalar matrices. We take them to be diagonal with  eigenvalues $(x_1,x_2)$ and $(y_1,y_2)$ respectively. Gauge invariance requires invariance under the simultaneous exchange of labels $1\leftrightarrow2$. We choose the four primaries
\begin{equation}
    p_1=\operatorname{Tr}X,\qquad p_2=\operatorname{Tr}X^2,
    \qquad q_1=\operatorname{Tr}Y,\qquad q_2=\operatorname{Tr}Y^2.
\end{equation}
The mixed trace \(S=\operatorname{Tr}(XY)=x_1y_1+x_2y_2\) is an additional invariant that cannot be expressed as a polynomial in these primaries.

Since $2S-p_1q_1=(x_1-x_2)(y_1-y_2)$, we obtain
\begin{equation}\label{eq:relation-1}
    (2S-p_1q_1)^2=(2p_2-p_1^2)(2q_2-q_1^2).
\end{equation}
Relation \eqref{eq:relation-1} reduces every power of $S$ to an expression at most linear in $S$. In fact, every invariant polynomial has the unique form
\begin{equation}
    \mathcal O=A(p_1,p_2,q_1,q_2)+S\,B(p_1,p_2,q_1,q_2),
\end{equation}
where $A$ and $B$ are polynomials.  This is argued as follows. Express each polynomial in the eigenvalue sums and differences. Simultaneous exchange changes the sign of both differences, so every invariant monomial has either two even exponents or two odd exponents. Terms with even exponents are polynomials in the primaries; those with odd exponents are polynomials in the primaries multiplied by $2S-p_1q_1$. If $A+SB=0$, exchanging $y_1$ and $y_2$ and subtracting gives $(x_1-x_2)(y_1-y_2)B=0$. Hence $B=0$, and then $A=0$.

Setting the four primaries to zero then gives the two-dimensional quotient algebra $\mathbb C[S]/(S^2)$.

\subsection{Counting primaries and secondaries}

Let $h_n$ count linearly independent operators containing $n$ scalar fields. Their Hilbert series is $H(t)=\sum_{n\geq0}h_nt^n$. 
Restricting to above simple models, a bosonic primary of degree $d$ can appear any number of times and therefore contributes the geometric series $1+t^d+t^{2d}+\cdots=(1-t^d)^{-1}$. For the one-matrix sector this gives
\begin{equation}
    H_N(t)=\prod_{k=1}^{N}\frac{1}{1-t^k}.
\end{equation}
For the two-matrix $U(2)$ example, the four primaries have degrees $1,2,1,2$, while the secondaries $1,S$ have degrees zero and two. Hence
\begin{equation}
    H_2(t)=\frac{1+t^2}{(1-t)^2(1-t^2)^2}.
\end{equation}
The denominator counts primary monomials; the numerator counts secondaries by degree.

Conversely, a known generating function can guide the construction of a Hironaka basis. For $m$ proposed bosonic primaries of positive degrees $d_i$, define
\begin{equation}
    P(t)=H(t)\prod_{i=1}^{m}(1-t^{d_i}).
\end{equation}
If these primaries admit a Hironaka decomposition, $P(t)$ is a polynomial whose coefficient of $t^j$ counts secondaries of degree $j$. Their total number is therefore
\begin{equation}
    r=P(1)=\lim_{t\to1}H(t)\prod_{i=1}^{m}(1-t^{d_i}).
\end{equation}
Moreover the order of the pole of $H(t)$ at $t=1$ gives the number $m$ of bosonic primaries. In the two-matrix example the pole has order four, and $P(t)=1+t^2$ gives two secondaries. Of course, a polynomial $P(t)$ with nonnegative integer coefficients is necessary but does not establish a Hironaka decomposition by itself. 

The primaries and secondaries may not have fixed flavor charges. We therefore use the unrefined Hilbert series, which counts operators only by field degree.

\subsection{The general operator expansion}

Choose $m$ algebraically independent bosonic primaries $p_i$ and $r$ secondaries $s_\alpha$. A Hironaka decomposition means that every operator can be written uniquely as
\begin{equation}
    \mathcal O=\sum_{\alpha=1}^{r}
    s_\alpha F_\alpha(p_1,\ldots,p_m),
\end{equation}
where the $F_\alpha$ are polynomials. The secondaries need not be independent under multiplication; instead, their products reduce to the same basis, with polynomial coefficients in the primaries, as $S^2$ did in the two-matrix example. These product relations tell us how operators multiply, information that the generating function alone does not contain.

Setting the primaries to zero gives a finite-dimensional quotient algebra whose vector-space basis consists of the images of the secondaries. If
\begin{equation}
    s_\alpha s_\beta
    =\sum_\gamma
    C_{\alpha\beta}^{\ \ \gamma}(p_1,\ldots,p_m)s_\gamma,
\end{equation}
then after the primaries are set to zero their multiplication becomes
\begin{equation}
    s_\alpha s_\beta
    =\sum_\gamma
    C_{\alpha\beta}^{\ \ \gamma}(0)s_\gamma.
\end{equation}

The separation into primaries and secondaries depends on which operators we choose as primaries. Even for a single unconstrained bosonic operator $x$, there are different choices. Taking $x$ as the primary leaves only the identity as a secondary. Taking $x^2$ as the primary instead gives two secondaries, $1$ and $x$, since every polynomial splits uniquely into even and odd powers:
\begin{equation}
    F(x)=A(x^2)+xB(x^2).
\end{equation}
The number of secondaries depends on the chosen primaries, even for the same operator space. It therefore does not count an independently defined class of physical states.

For a finitely generated graded commutative ring, admitting a Hironaka decomposition is equivalent to the Cohen–Macaulay property. Once this property is established, choose algebraically independent homogeneous primaries of positive degree such that setting them to zero leaves a finite-dimensional quotient. Any homogeneous basis of this quotient lifts to a secondary basis of the full ring \cite{Stanley:1979}. By the Hochster--Roberts theorem, the invariant ring of unconstrained complex bosonic fields under a linear action of a compact Lie group or finite group is Cohen--Macaulay and therefore admits a Hironaka decomposition~\cite{Hochster:1974}. This theorem guarantees the existence of Hironaka decompositions for the bosonic diagonal models and the ABJM scalar branch. We construct these decompositions below. 

The theorem is not applicable to the cohomology ring of a general supersymmetric theory. If every positive-degree scalar annihilates a nonzero class after multiplication, then no scalar primary acts freely. If the cohomology is also infinite-dimensional, finitely many secondaries cannot span it without primaries. Together these conditions rule out a Hironaka decomposition in such cases. Section~\ref{sec:other} gives a supersymmetric example of this failure.

In many examples, $P(t)$ is palindromic: degrees $j$ and $D-j$ contain equally many secondaries, where $D$ is the highest secondary degree. Because the identity is the sole degree-zero class, there is exactly one secondary in the top degree. For a commutative graded ring with a Hironaka decomposition, the Gorenstein property imposes a stronger condition: after setting the primary invariants to zero, every  nonzero homogeneous class must have a complementary-degree partner whose product is a nonzero multiple of the top-degree class. This implies a palindromic numerator. Conversely, for a commutative graded ring admitting a Hironaka decomposition, a palindromic numerator implies the Gorenstein property if the original ring has no zero divisors~\cite{Stanley:1978}.\footnote{A zero divisor is a nonzero element $a$ such that $ab=0$ for some nonzero $b$. In the two-matrix example, that quotient has $S\neq0$ but $S^2=0$.} Suppose a finite group acts linearly on complex bosonic fields, and no nonidentity element fixes a codimension-one subspace pointwise. Then the Hilbert-series numerator of the invariant ring is palindromic if and only if every group element has determinant one~\cite{Watanabe:1974a,Watanabe:1974b}. The Higgs branch of the class S theory corresponding to the four-punctured sphere has a symplectic singularity, so its coordinate ring is Gorenstein~\cite{Beauville_2000}. Its Hilbert-series numerator is therefore palindromic. we describe this ring in Section~\ref{sec:classs}.

\section{Commuting BPS sectors}
\label{sec:commuting}
\subsection{\texorpdfstring{The $1/4$-BPS model}{The 1/4-BPS model}}

The $1/4$-BPS model of $U(N)$ $\mathcal N=4$ super-Yang--Mills theory contains two scalar matrices $X,Y$ and one fermionic matrix $\Psi$.\footnote{We use the $1/4$-BPS label loosely, to parallel the $1/8$-BPS model below. The strict $1/4$-BPS count contains only $X,Y$, whereas the model studied here also retains $\Psi$  \cite[eqs.~(6.4) and (6.5)]{Kinney:2005ej}. This reference states that $\Psi$ is diagonal. This is justified when $X$ has distinct eigenvalues. Take the $Q$ action as $QX=QY=0$ and $Q\Psi=[X,Y]$. In an eigenbasis of $X$, with eigenvalues $x_i$, we have
\begin{equation*}
    Q\Psi_{ij}=(x_i-x_j)Y_{ij},\qquad QY_{ij}=0.
\end{equation*}
For $i\ne j$, the difference $x_i-x_j$ is invertible, so $Y_{ij}$ and $\Psi_{ij}/(x_i-x_j)$ form a $Q$-pair and are lifted from cohomology. Only the diagonal entries $x_i,y_i,\theta_i$ of $X,Y,\Psi$ remain, and gauge-invariance becomes invariance under simultaneous permutations of these diagonal entries.} Their diagonal components are $x_i,y_i,\theta_i$, with $i=1,\ldots,N$; the $\theta_i$ anticommute. Operators are polynomials in these components invariant under simultaneous permutations of the $N$ labels. Their grand canonical  generating function is given in equation (6.5) of~\cite{Kinney:2005ej}.

Let $u,v,w$ count the numbers of $X,Y,\Psi$, respectively. For $N=2$, the partition function is
\begin{equation}
    H^{1/4}_2(u,v,w)=
    \frac{(1+w)(1+uv+uw+vw)}
    {(1-u)(1-u^2)(1-v)(1-v^2)}.
\end{equation}
The denominator suggests retaining the four bosonic primaries $\operatorname{Tr}X,\operatorname{Tr}X^2,\operatorname{Tr}Y, \operatorname{Tr}Y^2$, while the factor $1+w$ accounts for the invariant fermionic sum $\eta=\operatorname{Tr}\Psi=\theta_1+\theta_2$, which can appear at most once. Once these dressings are separated, the remaining numerator predicts four secondaries: the identity, one operator containing an $X$ and a $Y$, and two fermionic operators containing either an $X$ or a $Y$ together with a $\Psi$. Secondaries: 
\begin{equation}
    1,\qquad S=\operatorname{Tr}(XY),\qquad
    U=\operatorname{Tr}(X\Psi),\qquad V=\operatorname{Tr}(Y\Psi).
\end{equation}
The weights are $1,uv,uw,vw$, as the numerator of the partition function requires.

To verify the basis, separate the sums from the differences:
\begin{equation}
    a=x_1-x_2,\qquad b=y_1-y_2,\qquad
    \delta=\theta_1-\theta_2.
\end{equation}
The simultaneous exchange of the two labels reverses all three differences and leaves $p_1,q_1,\eta$ unchanged. Since \(a^2=2p_2-p_1^2\), \(b^2=2q_2-q_1^2\), and \(\delta^2=0\), every invariant is a unique linear combination of
\[
1,\qquad ab,\qquad a\delta,\qquad b\delta,
\]
and their products with \(\eta\), with coefficients polynomial in the bosonic primaries. These four structures are equivalent to the proposed secondaries, since
\begin{equation}
    ab=2S-p_1q_1,\qquad a\delta=2U-p_1\eta,\qquad
    b\delta=2V-q_1\eta.
\end{equation}
This proves both spanning and independence. Multiplication of secondaries does not produce additional secondaries. For example,
\begin{equation}
    U^2=V^2=0,\qquad
    UV=\frac{\eta}{2}\bigl(p_1V-q_1U\bigr).
\end{equation}
Two fermionic secondaries can multiply to a nonzero operator, but their product is already contained in the same dressed basis.

For general $N$, we take the $2N$ bosonic primaries
\begin{equation}
    p_k=\operatorname{Tr}X^k,\qquad
    q_k=\operatorname{Tr}Y^k,\qquad k=1,\ldots,N,
\end{equation}
together with the fermionic dressing $\eta=\operatorname{Tr}\Psi$. The diagonal operator ring admits a Hironaka decomposition over these dressings (the general proof is given in Appendix~\ref{app:diagonal}), so its partition function has the form
\begin{equation}
    H_N^{1/4}(u,v,w)=
    \frac{(1+w)P_N(u,v,w)}
    {\prod_{k=1}^{N}(1-u^k)(1-v^k)},
\end{equation}
where $P_N$ counts the remaining secondaries.

Their total number can be extracted without computing the full numerator. We set $u=v=t$ and $w=1$, retaining the scalar degree but summing over all fermion numbers, and count gauge-invariant polynomials by averaging the action of all $N!$ permutations of the diagonal labels. The identity permutation contributes $2^N/(1-t)^{2N}$, since there are $2N$ independent bosonic components and $N$ fermionic components, each of which can appear at most once. Every nonidentity permutation has fewer independent cycles and so contributes a lower-order pole. At fixed $N$, then,
\begin{equation}
    H_N^{1/4}(t,t,1)\sim\frac{2^N}{N!(1-t)^{2N}},
    \qquad t\to1^-.
\end{equation}
Removing the primary factors and the two choices associated with $\eta$ gives
\begin{equation}
    r_N^{1/4}=P_N(1,1,1)=\frac12\lim_{t\to1^-}
    H_N^{1/4}(t,t,1)\prod_{k=1}^{N}(1-t^k)^2
    =N!\,2^{N-1}.
\end{equation}
This reproduces the four secondaries (without counting dressing by $\eta$) at $N=2$. At $N=3$ the numerator gives six secondaries with no fermions, twelve with one fermion and six with two fermions, for a total of twenty-four. These counts also agree with the oscillator partition function discussed in Section~\ref{sec:oscillator}. An explicit $N=3$ basis is listed in Appendix~\ref{app:u3}.

\subsection{\texorpdfstring{The $1/8$-BPS model}{The 1/8-BPS model}}

The $1/8$-BPS model contains three scalar matrices $X,Y,Z$ and two fermionic matrices $\Psi_1,\Psi_2$. Gauge invariance again requires simultaneous permutations of the diagonal labels. For $U(2)$ we choose the six bosonic primaries $\operatorname{Tr}X, \operatorname{Tr}X^2$, and the corresponding traces for $Y,Z$, together with the two fermionic dressings $\eta_a=\operatorname{Tr}\Psi_a$, $a=1,2$.

Let $u_1,u_2,u_3$ count the three scalar species and $w_1,w_2$ the two fermion species. The partition function is
\begin{equation}
    H_2^{1/8}=\frac{(1+w_1)(1+w_2)P_2^{1/8}}
    {\prod_{a=1}^{3}(1-u_a)(1-u_a^2)},
\end{equation}
where
\begin{align}
    P_2^{1/8}={}&(1+w_1w_2)(1+u_1u_2+u_2u_3+u_3u_1)\notag\\
    &+(w_1+w_2)(u_1+u_2+u_3+u_1u_2u_3).
\end{align}
Once the two fermion dressings are separated, this gives sixteen secondaries: four with no fermions, eight with one fermion and four with two fermions. Since the two-fermion secondaries are bosonic, the basis contains eight bosonic and eight fermionic operators, in agreement with the $N=2$, $(d,f)=(3,2)$ grand canonical oscillator partition function of Section~\ref{sec:oscillator}.


For $U(2)$, the basis follows from the exchange of the two diagonal entries. Denoting the scalar components by $x_i,y_i,z_i$ and the fermionic components by $\theta_{iA}$, where $i=1,2$ and $A=1,2$, we define the differences
\begin{equation}
    a=x_1-x_2,\qquad b=y_1-y_2,\qquad c=z_1-z_2,
    \qquad \delta_A=\theta_{1A}-\theta_{2A}.
\end{equation}
Each difference changes sign under the exchange, so a product is gauge invariant when it contains an even number of these factors. The squares of the scalar differences can be written in terms of the primaries, for example $a^2=2\operatorname{Tr}X^2-(\operatorname{Tr}X)^2$, while the fermionic differences square to zero. We can thus choose the secondaries as
\begin{equation}
\begin{array}{ll}
\text{No fermions:}&1,\ ab,\ bc,\ ca,\\[2pt]
\text{One fermion:}&a\delta_A,\ b\delta_A,\ c\delta_A,\ abc\delta_A,
\quad A=1,2,\\[2pt]
\text{Two fermions:}&\delta_1\delta_2,\ ab\delta_1\delta_2,
\ bc\delta_1\delta_2,\ ca\delta_1\delta_2.
\end{array}
\end{equation}
These are precisely the sixteen operators counted by the numerator. The connection to mixed traces is explicit:
$ab=2\operatorname{Tr}(XY)-\operatorname{Tr}X\,\operatorname{Tr}Y$,
with analogous expressions for the other pairs.


\subsection{The general count}

The same construction applies to a diagonal model with $N$ labels,
$d\geq1$ bosonic species and $f$ fermionic species, with no relations
other than fermion anticommutation and simultaneous permutation
invariance. For each bosonic species we choose the first $N$ traces as
primaries, and for each fermionic species we separate the sum of its
diagonal components, which gives $dN$ bosonic primaries and $f$ fermionic
dressings. We prove in Appendix~\ref{app:diagonal} that the remaining operators admit a
finite secondary basis.

If $t$ counts the total bosonic degree and $w$ the total fermion number, then
\begin{equation}
    H_{N;d,f}(t,w)=\frac{(1+w)^f P_{N;d,f}(t,w)}
    {\prod_{k=1}^{N}(1-t^k)^d}.
\end{equation}
As in the $1/4$-BPS model, the identity permutation gives the leading
pole as $t\to1^-$, with contribution
\begin{equation}
    H_{N;d,f}(t,w)\sim\frac{(1+w)^{Nf}}{N!(1-t)^{dN}}.
\end{equation}
Comparing the two expressions gives
\begin{equation}
    P_{N;d,f}(1,w)=(N!)^{d-1}(1+w)^{f(N-1)}.
\end{equation}
The coefficient of $w^F$ counts secondaries containing $F$
fermions, summed over all bosonic degrees, and setting $w=1$ gives their
total number,
\begin{equation}
    r_{N;d,f}=(N!)^{d-1}2^{f(N-1)}.
\end{equation}
Note that here we have separated $f$ fermionic dressings in the secondary counting.

For the $1/8$-BPS model, $d=3$ and $f=2$, so the number is
$(N!)^2 4^{N-1}$. At $N=2$, the fermion-counting polynomial is
$4(1+w)^2$, reproducing the four, eight and four operators constructed
above. The next
subsection explains this count in terms of the grand-canonical
partition function of the corresponding oscillator problem.

\subsection{Grand-canonical and oscillator interpretation}
\label{sec:oscillator}

The diagonal model has a simple oscillator interpretation. Each diagonal
label may be viewed as a particle carrying $d$ bosonic and $f$ fermionic
oscillator modes. 
A state of the $N$-particle system is then a polynomial in the $dN$ bosonic creation operators and the $fN$ fermionic creation operators, invariant under simultaneous permutations of the $N$ particle labels, which is precisely the diagonal operator space studied above. The oscillator description reproduces the $1/4$-BPS and $1/8$-BPS counting functions in equations (6.5) and (6.3), respectively, of~\cite{Kinney:2005ej}.

Let $t$ count bosonic excitations and $w$ count fermionic excitations.
For one particle, the bosonic oscillators give arbitrary occupation
numbers, while each fermionic oscillator may be occupied at most once,
so the one-particle states are labelled by $d$ nonnegative integers
and by a choice of occupied fermionic modes. A state with an even number
of occupied fermionic modes is bosonic, and one with an odd number is
fermionic.

It is useful to collect systems with different numbers of particles by
introducing $p$ to count $N$. For each one-particle state of weight
$t^n w^F$, Bose statistics gives a factor
\begin{equation}
    \frac{1}{1-p\,t^n w^F},
\end{equation}
when $F$ is even, while Fermi statistics gives
\begin{equation}
    1+p\,t^n w^F,
\end{equation}
when $F$ is odd. The grand canonical partition function is obtained by the plethystic exponent of the single-particle partition function.
\begin{equation}
    \sum_{N=0}^{\infty}p^N H_{N;d,f}(t,w)
    =
    \exp\left[
    \sum_{\ell=1}^{\infty}\frac{p^\ell}{\ell}
    \frac{\bigl[1-(-w)^\ell\bigr]^f}
    {(1-t^\ell)^d}
    \right].
\end{equation}
Setting $(d,f)=(2,1)$ and $(3,2)$ gives the grand-canonical counting
functions of the $1/4$-BPS and $1/8$-BPS models, respectively.

The secondary counts follow from a simple limit. Replacing \(p\) by \(z(1-t)^d\) and taking \(t\to1^-\) at each fixed power of \(z\), we find
\begin{equation*}
    \lim_{t\to1^-}
    \sum_{N=0}^{\infty}
    \bigl[z(1-t)^d\bigr]^N H_{N;d,f}(t,w)
    =\exp\bigl[z(1+w)^f\bigr].
\end{equation*}
The coefficient of \(z^N\), \((1+w)^{Nf}/N!\), gives the leading-pole coefficient at fixed \(N\). Multiplying by \((N!)^d\) gives the secondary-counting polynomial \((N!)^{d-1}(1+w)^{Nf}\) over the bosonic primaries. Separating the \(f\) fermionic dressings further divides this polynomial by \((1+w)^f\).


\section{Other supersymmetric operator rings}
\label{sec:other}
\subsection{A Wess--Zumino model}

The diagonal models above involve free scalar components, subject only to permutation invariance. In other supersymmetric sectors, the supercharge imposes relations among scalar products, as in the usual chiral-ring description~\cite{Cachazo:2002ry}. We first examine a simple model in which these relations still allow a Hironaka decomposition.

Consider a Wess--Zumino model with three chiral multiplets, scalar components $x,y,z$, and superpotential $W=xyz$. The corresponding fermionic variables $\psi_x,\psi_y,\psi_z$ obey
\begin{equation}
    Qx=Qy=Qz=0,\qquad
    Q\psi_x=yz,\qquad Q\psi_y=xz,\qquad Q\psi_z=xy.
\end{equation}
We will only focus on cohomology in the polynomial (non-derivative) sector.  The mixed scalar products $xy,xz,yz$  vanish in cohomology, since each is $Q$ acting on another operator, while powers of a single scalar survive. The scalar part of the operator space thus consists of three towers, built from $x$, $y$, and $z$, sharing the identity.

Assign degree one to each scalar and degree two to each fermion, so that \(Q\) preserves the degree. Computing the cohomology at each fermion number using Macaulay2 gives the Hilbert series,
\begin{equation}
    H_0(t)=\frac{1+2t}{1-t},\qquad H_1(t)=\frac{2t^3+t^4}{1-t}\qquad H_2(t)= H_3(t)=0.
\end{equation}
This suggests one bosonic primary of degree one and six secondaries. We will show that $p=x+y+z$ is a suitable primary: every cohomology class has a unique expansion in six fixed operators multiplied by polynomials in $p$.

The three scalar secondaries can be chosen as $1,y,z$. Indeed, $x=p-y-z$, while the vanishing mixed products imply $y^2=py$ and $z^2=pz$. All higher scalar powers therefore reduce to these three operators multiplied by powers of $p$.

For the fermionic operators, define
\begin{equation}
    a=x\psi_x-y\psi_y,\qquad
    b=y\psi_y-z\psi_z,\qquad c=ya.
\end{equation}
Both $a$ and $b$ are $Q$-closed because the two terms in each expression have the same $Q$-variation, $xyz$, and together they generate the one-fermion cohomology under scalar multiplication. Acting with $Q$ on two-fermion products gives further relations,
\begin{equation}
    za=0,\qquad xb=0,\qquad y(a+b)=0
    \quad\text{in cohomology}.
\end{equation}
For example, $Q(\psi_x\psi_y)=-za$. These relations reduce scalar multiples of $a,b,c$ to the same three operators:
\begin{equation}
\begin{aligned}
    xa&=pa-c,& ya&=c,& za&=0,\\
    xb&=0,& yb&=-c,& zb&=pb+c,\\
    xc&=0,& yc&=pc,& zc&=0.
\end{aligned}
\end{equation}
There are no surviving classes with two or more fermions.

Thus every class is a sum of the six secondaries
\begin{equation}
    1,\qquad y,\qquad z,\qquad a,\qquad b,\qquad c,
\end{equation}
multiplied by polynomials in $p$. Their degrees are $0,1,1,3,3,4$, which reproduce the numerator of $H(t)$. Since they span the cohomology and give exactly its number of states at each degree, the expansion is unique. 

\subsection{When Hironaka decomposition fails}

Now consider two chiral multiplets with scalar components $x,y$ and superpotential $W=x^2y^2$. In the same polynomial sector, the supercharge acts as
\begin{equation}
    Qx=Qy=0,\qquad Q\psi_x=2xy^2,\qquad Q\psi_y=2x^2y.
\end{equation}
The operator $xy$ is $Q$-closed and is not $Q$-exact: every exact scalar polynomial contains either $xy^2$ or $x^2y$ as a factor. Yet multiplying $xy$ by either scalar gives an exact operator,
\begin{equation}
    x(xy)=\tfrac12 Q\psi_y,\qquad
    y(xy)=\tfrac12 Q\psi_x.
\end{equation}
So $xy$ survives in cohomology, yet every scalar dressing of positive degree kills it.

This prevents a Hironaka decomposition over any choice of positive-degree scalar primaries. In such a decomposition, multiplying a nonzero class by a primary cannot give zero: its polynomial coefficients would simply acquire another factor of that primary. The class $xy$ violates this requirement for every possible scalar choice. Since the powers $x^n$ remain nonzero in cohomology, this is the failure described in Section~\ref{sec:hironaka}: the cohomology is not Cohen--Macaulay. Unlike the previous model, the spectrum cannot be organized entirely into independent towers generated by scalar multiplication.

The generating function also signals the problem. Assigning degree one to the scalars and degree three to the fermions gives
\begin{equation}
    H(t)=\frac{1+t+t^2-t^3+t^4+t^5}{1-t}.
\end{equation}
The negative coefficient in the numerator rules out interpreting this expression as a secondary count for a degree-one primary. The operator argument also excludes scalar primaries of higher degree.

Interestingly the counting function can also be written as
\begin{equation}
    H(t)=\frac{1}{1-t}+\frac{t}{1-t}+t^2 +\frac{t^4}{1-t}+\frac{t^5}{1-t}.
\end{equation}
These terms count the classes
\[
x^n,\qquad y^{n+1},\qquad xy,\qquad
sx^n,\qquad sy^{n+1},\qquad n\geq0,
\]
where \(s=x\psi_x-y\psi_y\) is a one-fermion cohomology class of degree four. The class \(xy\) has no scalar dressings, while the other classes allow dressing form towers generated by \(x\) (or \(y\)). This organization is a Stanley decomposition: different seed classes have different polynomial dressings, while the expansion remains unique. It accommodates the isolated class \(xy\) alongside the infinite towers. Stanley-type decompositions may therefore be flexible enough to organize tree-level cohomology in general supersymmetric field theories. We do not pursue this possibility here.

\subsection{An $\mathcal N=1$ gauge theory}
\label{sec:kw}

We next ask whether scalar relations and a nonabelian gauge projection can still have a Hironaka decomposition. Consider the Klebanov--Witten theory~\cite{Klebanov_1998} with gauge group $SU(2)\times SU(2)$, two bifundamental chiral fields $A_1,A_2$ and two fields $B_1,B_2$ in the conjugate representation. We study its scalar F-term ring, with superpotential
\begin{equation}
 W=\operatorname{Tr}(A_1B_1A_2B_2-A_1B_2A_2B_1).
\end{equation}
The F-terms are the matrix derivatives of $W$. At rank two, the epsilon tensor identifies the fundamental and antifundamental representations of each \(SU(2)\). We can therefore write all four fields in the same \((\mathbf2,\mathbf2)\) representation. The pair of indices \((a,\dot a)\), each taking two values, forms a single four-component vector index of \(SO(4)\). Denoting these vectors by \(v_i\), the superpotential becomes \(W\propto\det(v_1,v_2,v_3,v_4)\). Its derivatives then set all three-by-three minors of the matrix of vectors to zero.

Consequently, the scalar cohomology is given by the ten Gram entries $g_{ij}=v_i\cdot v_j$, where the dot denotes the contraction in \(SO(4)\), subject to the \(3\times3\) minors of the symmetric matrix $G=(g_{ij})$ vanishing. This means that \(G\) has rank at most two. We can therefore write \(G=VV^{\mathsf T}\), where the \(i\)th row of \(V\) is \((x_i,y_i)\in\mathbb C^2\), with \(i=1,\ldots,4\). The matrix \(G\) is unchanged under \(V\mapsto VO\), with \(O\in O(2)\). Its entries generate all polynomial invariants of this action. We therefore obtain the Hilbert series by counting \(O(2)\)-invariant polynomials in \(x_i,y_i\).

Introduce \(a_i=x_i+\mathrm{i}y_i\) and \(b_i=x_i-\mathrm{i}y_i\). Rotations act on \(a_i\) and \(b_i\) with opposite phases, so a rotation-invariant monomial must contain equal numbers of \(a\)'s and \(b\)'s. There are
\[
d_n=\binom{n+3}{3}
\]
degree-\(n\) monomials in each set. A reflection exchanges the two sets, leaving one invariant combination for each unordered pair of such monomials. Thus
\[
h_{2n}=\frac{d_n(d_n+1)}{2},\qquad h_{2n+1}=0.
\]
Summing these contributions gives
\begin{equation}\label{eq:kw-series}
H_{\mathrm{KW}}(t)
=\sum_{n=0}^{\infty}\frac{d_n(d_n+1)}{2}\,t^{2n}
=\frac{1+3t^2+6t^4}{(1-t^2)^7}=1+10t^2+55t^4+210t^6+\cdots.
\end{equation}
This agrees with the $SU(2)\times SU(2)$ Klebanov--Witten Hilbert series appearing in the integrand eq. 4.25 of \cite{Cremonesi:2016nbo}, after setting the baryonic fugacity $b$ to one.

Each elementary scalar is taken to have degree one. The numerator is not palindromic: this ring has a decomposition but not the symmetry discussed in Section~\ref{sec:hironaka}. In Appendix~\ref{app:kw}, we give seven degree-two primaries and show that, after setting them to zero, the remaining scalar operators are spanned by
\begin{equation}
    1,\qquad x,y,z,\qquad x^2,xy,xz,y^2,yz,z^2,
\end{equation}
where 
\begin{equation*}
    x=g_{33}=v_3\cdot v_3,\qquad y=g_{34}=v_3\cdot v_4,\qquad z=g_{44}=v_4\cdot v_4.
\end{equation*}
To express the basis in the original bifundamental fields, identify
\[
M_1=A_1,\qquad M_2=A_2,\qquad
M_3=\operatorname{adj}B_1,\qquad M_4=\operatorname{adj}B_2,
\]
where \(\operatorname{adj}\) denotes the adjugate matrix. With the normalization
\[
g_{ij}=\frac12\operatorname{Tr}\!\left(M_i\operatorname{adj}M_j\right),
\]
the three secondary generators become
\[
x=\det B_1,\qquad
y=\frac12\operatorname{Tr}\!\left(B_1\operatorname{adj}B_2\right),
\qquad z=\det B_2.
\]
The same identification expresses the seven primaries of Appendix F.2 in the original fields.

The chosen primaries are homogeneous in field degree but are not chosen to have definite values of all flavor charges. This illustrates the point made in Section~\ref{sec:hironaka}: the Hironaka decomposition need not be visible at each fixed flavor charge, even though it has a simple unrefined description. We show in Appendix~\ref{app:kw} that the ten operators above lift to a Hironaka secondary basis of the full scalar ring. 

\subsection{Higgs branches of class-S theories}
\label{sec:classs}

We consider theories of class S, obtained from the six-dimensional $(2,0)$ theory on a punctured Riemann surface. For the $A_1$ theory on a sphere with full punctures, the Hall--Littlewood limit of the superconformal index equals the Higgs-branch Hilbert series~\cite{Gadde:2011uv}. This gives $H^0(t)$ in this family.

The four-punctured sphere gives $SU(2)$ gauge theory with four fundamental hypermultiplets. Combine their scalar fields into eight doublets $q_i^A$, with $i=1,2$ and $A=1,\ldots,8$. The flavor moment maps
\begin{equation}
 m_{AB}=\epsilon^{ij}q_i^Aq_j^B=-m_{BA},\qquad A<B,
\end{equation}
are the 28 quadratic generators of the Higgs-branch ring; we use $\epsilon^{12}=1$. Their relations describe the closure of the minimal nilpotent orbit of $SO(8)$, or equivalently the centered one-instanton moduli space~\cite{Benvenuti_2010}. The series is
\begin{equation}
\label{eq:four-puncture-series}
 H_4(t)=\frac{1+18t^2+65t^4+65t^6+18t^8+t^{10}}{(1-t^2)^{10}}.
\end{equation}
We construct ten degree-two primaries and all 168 secondaries in Appendix~\ref{app:classs}. The chosen primaries are homogeneous in field degree but are linear combinations of moment maps with different flavor weights, so the decomposition is naturally visible in the unrefined series rather than as a decomposition at each fixed flavor charge. After setting the primaries to zero, exact reduction by the relations verifies the proposed basis. The Cohen--Macaulay property then lifts this basis to a decomposition in every degree (Section~\ref{sec:hironaka}).

Appendix~\ref{app:more-punctures} computes the numerators for five-, six- and seven-punctured spheres. For suitable denominator choices up to twenty punctures, we find positive palindromic numerators. This symmetry is consistent with the corresponding Higgs-branch rings being Gorenstein (Section~\ref{sec:hironaka}). For the four-puncture case, we verify the symmetry explicitly by constructing a Hironaka basis (Appendix~\ref{app:classs}).

The class S theory corresponding to a closed genus-two surface also admits a Hironaka decomposition. It has two primaries and two secondaries. For surfaces with nonzero genus, the HL index does not equal the Hilbert series of \(H^0\), because nonzero higher cohomology involving fermions also contributes to the index~\cite{Gadde:2011uv}. 
In its dumbbell gauge-theory frame, the scalar ring has three generators \(M,B_1,B_2\) of degrees \(2,4,4\) and, with the normalization specified in Appendix~\ref{app:genus-two}, one relation:
\begin{equation}
\label{eq:genus-two-ring}
B_2^2+B_1^2+2M^2B_1=0,\qquad
H_{g=2}(t)=\frac{1+t^4}{(1-t^2)(1-t^4)}.
\end{equation}
This presentation follows from the Higgs-branch calculation of~\cite{Hanany_2011}. The coefficient of $B_2^2$ is one, so polynomial division reduces every scalar operator uniquely to $F(M,B_1)+B_2G(M,B_1)$. The secondaries over the primaries $M,B_1$ are therefore $1,B_2$. 

\subsection{The scalar cohomology of ABJM theory}

The scalar cohomology ring of ABJM theory admits a Hironaka decomposition for every $N$ and positive integer $k$. The theory has gauge group $U(N)_k\times U(N)_{-k}$, and describes $N$ identical M2-branes on $\mathbb C^4/\mathbb Z_k$. We study the ring of holomorphic functions on this branch. This includes matter operators and gauge-invariant dressed monopole operators, which contain insertions of quantized magnetic flux~\cite{Kim:2009wb,Cremonesi:2016nbo}.

For one brane, choose the four complex coordinates $A_1,A_2,B_1,B_2$ of the usual chiral description~\cite{Aharony:2008ug,Cremonesi:2016nbo}. The discrete identification is generated by
\begin{equation}
    (A_1,A_2,B_1,B_2)\longmapsto
    (\omega A_1,\omega A_2,\omega^{-1}B_1,\omega^{-1}B_2),
    \qquad \omega=e^{2\pi i/k}.
\end{equation}
The opposite phases of the $A$ and $B$ coordinates determine which products survive: a monomial $A_1^{n_1}A_2^{n_2}B_1^{n_3}B_2^{n_4}$ is invariant precisely when $n_1+n_2-n_3-n_4$ is a multiple of $k$.

We may choose $A_1^k,A_2^k,B_1^k,B_2^k$ as primaries. From any invariant monomial, extract as many powers of these four operators as possible; the remaining exponents then lie between zero and $k-1$ and still satisfy the invariance condition. These remaining monomials form a secondary basis, and the separation is unique because division of each exponent by $k$ has a unique quotient and remainder. There are $k^3$ secondaries, since any choice of three remaining exponents fixes the fourth modulo $k$. The decomposition thus follows directly from the discrete identification.

Let $t$ count the total degree in the four brane coordinates. We can count invariant monomials by averaging over the $k$ discrete phases:
\begin{equation}
    H_{1,k}(t)=\frac1k\sum_{j=0}^{k-1}
    \frac{1}{(1-\omega^jt)^2(1-\omega^{-j}t)^2}.
\end{equation}
Expanding each denominator generates all monomials with their phase weights, and the sum over $j$ removes those that are not invariant. It takes the form,
\begin{equation}
    H_{1,k}(t)=\frac{P_{1,k}(t)}{(1-t^k)^4}.
\end{equation}
Here $P_{1,k}(t)$ counts the invariant monomials whose four exponents are each less than $k$, so $P_{1,k}(1)=k^3$. The highest power of $t$ is $4(k-1)$.

For example, at $k=3$ we find
\begin{equation}
    P_{1,3}(t)=1+4t^2+4t^3+9t^4+4t^5+4t^6+t^8.
\end{equation}
The four degree-two secondaries are $A_iB_j$, with $i,j=1,2$, since the opposite phases cancel in each product, and the coefficients sum to twenty-seven, as expected. The numerator is palindromic, as the criterion of Section~\ref{sec:hironaka} requires, because the phases $(\omega,\omega,\omega^{-1},\omega^{-1})$ have determinant one. 

The Hilbert series for \(N\) branes, encoded into a grand canonical partition function, is obtained by plethystic exponential of a single brane Hilbert series. Introducing \(p\) to count branes, we obtain~\cite{Cremonesi:2016nbo}
\[
\mathcal Z_k(p,t)
=\sum_{N=0}^{\infty}p^N H_{N,k}(t)
=\operatorname{PE}\!\left[pH_{1,k}(t)\right]
=\exp\!\left[
\sum_{\ell=1}^{\infty}\frac{p^\ell}{\ell}
H_{1,k}(t^\ell)
\right].
\]
The coefficient of \(p^N\) gives the fixed-\(N\) Hilbert series. At \(k=1\), this reduces to the grand-canonical counting function for four bosonic species with only permutation invariance.

Write \(z_i^{(a)}\) for coordinate \(a=1,\ldots,4\) of brane \(i=1,\ldots,N\). A convenient choice of primaries is
\[
p_r^{(a)}=\sum_{i=1}^{N}\bigl(z_i^{(a)}\bigr)^{kr},
\qquad r=1,\ldots,N.
\]
Each term is invariant under the discrete phases, and the sum is invariant under brane exchange. There are \(4N\) independent primaries, with four of each degree \(k,2k,\ldots,Nk\). The scalar ring admits a Hironaka decomposition over these primaries; the proof is given in Appendix~\ref{app:abjm}. Thus
\[
H_{N,k}(t)=\frac{P_{N,k}(t)}
{\prod_{r=1}^{N}(1-t^{kr})^4},
\]
where \(P_{N,k}\) counts the secondaries.

Their total number follows from the same grand-canonical limit used for the diagonal models. Since
\[
H_{1,k}(t)\sim\frac{1}{k(1-t)^4},
\qquad t\to1^-,
\]
replacing \(p\) by \(\xi(1-t)^4\) suppresses every term with \(\ell>1\) in the exponent. Here \(\ell\) labels the cycle length in permutations of the branes. Taking the limit at each fixed power of \(\xi\), we find
\[
\lim_{t\to1^-}
\sum_{N=0}^{\infty}
\bigl[\xi(1-t)^4\bigr]^N H_{N,k}(t)
=\exp(\xi/k).
\]
The coefficient of \(\xi^N\) therefore gives
\[
H_{N,k}(t)\sim\frac{1}{k^N N!(1-t)^{4N}}.
\]
Multiplying by the primary factors gives the total secondary count,
\begin{equation}\label{eq:ABJMsc_count}
    r_{N,k}=P_{N,k}(1)
=\frac{\prod_{r=1}^{N}(kr)^4}{k^N N!}
=k^{3N}(N!)^3.
\end{equation}

For one brane this reduces to \(k^3\), as found directly, while at \(k=1\) it gives \((N!)^3\), the result for four bosonic species. Brane exchanges and the discrete phases act on \(\mathbb C^{4N}\) with determinant one and fix no codimension-one subspace, so \(P_{N,k}\) is palindromic for every \(N\) and \(k\).

\subsection{A restricted bifundamental complex}
\label{sec:restricted}

To include fermions with a nonabelian gauge projection, we consider a polynomial complex motivated by the cubic supersymmetry variations in ABJM theory. Let $A,C,\psi$ transform as bifundamentals of $SU(2)\times SU(2)$ and let $B$ transform in the conjugate representation. The matrices $A,B,C$ are bosonic and $\psi$ is Grassmann odd~\cite{Kim:2009wb}. The differential is
\begin{equation}
 QA=QB=QC=0,\qquad Q\psi=ABC-CBA.
\end{equation}
The epsilon tensors identify the conjugate bifundamental \(B\) with a bifundamental, so the three scalars become vectors \(v_1,v_2,v_3\) of \(SO(4)\). After a constant normalization of \(\psi\), the differential becomes
\begin{equation}
\label{eq:restricted-Q}
Qv_i^\mu=0,\qquad Q\psi_\mu=T_\mu,\qquad
T_\mu=\epsilon_{\nu\rho\sigma\mu}\,
v_1^\nu v_2^\rho v_3^\sigma.
\end{equation}
Here \(i=1,2,3\), the vector indices run from one to four, and \(\epsilon_{1234}=1\). Taking polynomial invariants under \(SO(4)\) imposes the \(SU(2)\times SU(2)\) gauge projection.

Assign degree one to each scalar and degree three to $\psi$, so that $Q$ preserves the grading. Let $\mathcal H_j$ denote cohomology at fermion number $j$, and $H_j(t)$ its Hilbert series. We find
\begin{equation}
\label{eq:restricted-series}
 H_0(t)=\frac{1+t^2+t^4}{(1-t^2)^5},\qquad
 H_1(t)=\frac{3t^4+3t^6}{(1-t^2)^5},\qquad
 H_2(t)=\frac{3t^8}{(1-t^2)^5},
\end{equation}
with $\mathcal H_j=0$ for $j\geq3$. Appendix~\ref{app:restricted} constructs a common set of five degree-two scalar primaries and proves freeness, with three scalar, six one-fermion and three two-fermion secondaries. This example effectively extends the Hironaka decomposition of Wess--Zumino models to a nonabelian invariant complex.


\section{\texorpdfstring{Graviton dressings and finite-$N$ states}{Graviton dressings and finite-N states}}
\label{sec:graviton}

A physical application of Hironaka decomposition is to ask how BPS states can be dressed by gravitons. On the gauge-theory side, graviton dressing means multiplying a cohomology class by an operator dual to a graviton. The product is $Q$-closed, but it need not give a new state: it may be exact or related to other products. A Hironaka decomposition would identify a set of dressings that can be applied independently, together with the operators on which they act.

We study this question in the BMN truncation of $\mathcal N=4$ super-Yang--Mills theory~\cite{Choi:2023znd,Choi:2023vdm,Gadde:2025yoa}. It contains three adjoint scalars $X,Y,Z$, three adjoint fermions $\psi_1,\psi_2,\psi_3$, and one field-strength component $F$, without derivatives. We use the tree-level action of the supercharge~\cite{Grant:2008sk,Chang:2013fba},
\begin{equation}
\begin{aligned}
    QX&=QY=QZ=0,\\
    Q\psi_1&=[Y,Z],\qquad Q\psi_2=[Z,X],\qquad Q\psi_3=[X,Y]\\
    QF&=[X,\psi_1]+[Y,\psi_2]+[Z,\psi_3]
\end{aligned},
\end{equation}
The cohomology is gauge-invariant operators subject to these relations.

Unlike the diagonal models of Section~\ref{sec:commuting}, this problem has noncommuting matrices and a nontrivial action of $Q$. We first consider graviton operators and set the fermions to zero. They are constructed from bosonic matrices that are diagonal. The resulting ring admits an explicit decomposition. We also discuss its extension to the full cohomology.

\subsection{The bosonic part of the diagonal graviton ring}

For $SU(2)$, the graviton operators in the BMN sector are generated by the scalar bilinears $\operatorname{Tr}(\phi^m\phi^n)$, with $\phi^m=(X,Y,Z)$, and their supersymmetry descendants. We use their diagonal realization described in~\cite{Choi:2023znd,Gadde:2025yoa}. Write the scalar eigenvalues as $(x,-x)$, $(y,-y)$ and $(z,-z)$, and the field-strength eigenvalues as $(f,-f)$. Evaluating the graviton generators on these letters and setting the fermionic components to zero gives the ring generated by
\begin{equation}
    x^2,\quad y^2,\quad z^2,\quad xy,\quad yz,\quad zx,
    \quad xf,\quad yf,\quad zf.
\end{equation}
These denote the bosonic projections of supersymmetric operators, not purely bosonic representatives: for example, $\operatorname{Tr}(XF)$ generally needs additional fermion terms to be $Q$-closed.

With degree two assigned to $x,y,z$ and degree six to $f$, the Hilbert series of this ring is
\begin{equation}
    H_{\mathrm{grav}}^{(0)}(t)=
    \frac{1+5t^4+9t^8+5t^{12}+t^{16}}
    {(1-t^4)(1-t^8)^3}.
\end{equation}
We construct the primaries and secondaries suggested by this formula in  Appendix~\ref{app:bmn}, where we also show that a decomposition of this ring must exist. Multiplying any proposed secondary by a ring generator reduces to the same list, with polynomial coefficients in the primaries. Since the list contains the identity, it spans the ring. Its dressed generating function equals the Hilbert series of the ring, so the expansion is unique. This proves the Hironaka decomposition.

\subsection{Dressing operators outside the graviton sector}

Explicit operators outside the graviton sector have been constructed at low rank~\cite{Chang:2022mjp,Choi:2022caq,Choi:2023znd,Choi:2023vdm}. The $SU(2)$ BMN index provides an interesting example of graviton dressing. Let $Z(t)$ be the full index and $Z_{\mathrm{grav}}(t)$ its graviton contribution, with scalars, fermions and the field strength assigned weights $t^2,t^4,t^6$, respectively. Their difference is~\cite{Choi:2023znd}
\begin{equation}
    Z(t)-Z_{\mathrm{grav}}(t)
    =-\frac{t^{24}}{1-t^{12}}\,
    \frac{(1-t^2)^3}{(1-t^8)^3}.
\end{equation}
Choi et al.~\cite{Choi:2023znd} construct an infinite sequence of fermionic classes $\mathcal O_n$ of degree $24+12n$, with $n\geq0$. Their contribution gives the factor $-t^{24}/(1-t^{12})$. Three fermionic supersymmetry generators acting on these classes, each raising the degree by two, account for $(1-t^2)^3$. Dressings by three bosonic graviton operators of degree eight account for the remaining factor $(1-t^8)^{-3}$. Some products with other graviton operators are $Q$-exact and therefore vanish in cohomology.

This suggests asking whether the surviving operators have a unique expansion under the allowed dressings. The index alone cannot answer this question, because bosonic and fermionic contributions can cancel. Nor does the factor $1/(1-t^{12})$ by itself identify a bosonic operator whose repeated multiplication generates that sequence. To establish a Hironaka decomposition, we would need the operators themselves and their multiplication relations.

The Wess--Zumino counterexample explains why the choice of dressings is important. If a nonzero class becomes exact after multiplication by a graviton operator, that graviton cannot serve as an unrestricted primary for a space containing the class. The appropriate question is therefore whether a smaller set of graviton dressings acts independently and leaves a finite secondary basis. We leave this question open.

\subsection{\texorpdfstring{Finite-$N$ relations and fortuity}{Finite-N relations and fortuity}}

Trace relations can do more than identify products of BPS operators: they can also make an operator $Q$-closed. To see how, first consider trace expressions before imposing identities specific to a matrix size, following the covering construction of~\cite{Chang:2024zqi}. Suppose an expression $\widetilde{\mathcal O}$ obeys
\begin{equation}
    Q\widetilde{\mathcal O}=R_N,
\end{equation}
where $R_N$ is a trace expression that vanishes for $N\times N$ matrices. After imposing this identity, the resulting operator $\mathcal O$ is $Q$-closed, and it defines a nonzero cohomology class if it is not $Q$-exact.

Fortuity means that a cohomology class owes its existence to finite-\(N\) trace identities. These identities make the operator \(Q\)-closed, and no alternative expression for the same class is closed without using them~\cite{Chang:2024zqi}. Checking a single expression is therefore insufficient: the class might have another expression that is closed at every rank.

A Hironaka decomposition and fortuity answer different questions. The decomposition tells us how to build operators by multiplication at a fixed $N$, whereas fortuity asks which classes require the trace relations at that $N$ in order to exist. The two can be studied together when the same dressings remain available as $N$ changes: additional classes may then appear among the secondaries. A simple model illustrates this possibility.

Take two bosonic variables $x,z$ and one fermionic variable $\psi$, with
\begin{equation}
    Qx=Qz=0,\qquad Q\psi=x^2.
\end{equation}
To model a rank-dependent trace relation we impose $x^{N+1}=0$, with $N\geq1$; this is a toy model and does not follow from any gauge theory. The variable $z$ is unrestricted and will play the role of a dressing available at every $N$.

Without the rank-dependent relation, the bosonic cohomology consists of $1,x$ multiplied by arbitrary polynomials in $z$, since $x^2$ is exact. There are no fermionic classes: for a polynomial $f(x,z)$, the condition $Q(f\psi)=fx^2=0$ requires $f=0$. After imposing $x^{N+1}=0$, however, two fermionic operators become closed,
\begin{equation}
    Q(x^{N-1}\psi)=x^{N+1}=0,\qquad
    Q(x^N\psi)=x^{N+2}=0.
\end{equation}
Neither is exact, because there is no two-fermion operator whose $Q$-variation could produce it. Every cohomology class now has the unique form
\begin{equation}
    \mathcal O=A(z)+xB(z)+x^{N-1}\psi\,C(z)+x^N\psi\,D(z),
\end{equation}
where $A,B,C,D$ are polynomials.

Thus $z$ remains an unrestricted primary, while the two new fermionic secondaries are fortuitous. Their existence depends on the rank-dependent relation, but their dressing by $z$ does not. It would be interesting to look for a similar structure in the BMN cohomology. 

\section{Summary and outlook}
\label{sec:discussion}

We explore whether tree-level supersymmetric cohomology admits a Hironaka decomposition. We prove this for diagonal models with bosons and fermions and construct secondary bases for several interacting examples, including Wess–Zumino, Klebanov–Witten, class-S and ABJM scalar rings, and a nonabelian complex containing fermions. For BMN, we establish the decomposition only for the bosonic diagonal image of the graviton ring. The WZ model with superpotential \(W=x^2y^2\) provides a counterexample to Hironaka decomposition. Our work gives rise to the following interesting questions.

\begin{itemize}
\item
Which supersymmetric theories admit a Hironaka decomposition of their tree-level cohomology? Can we identify conditions on the supersymmetry relations and finite-rank matrix identities that ensure its existence? The $W=x^2y^2$ example fails to admit such a decomposition but admits a Stanley decomposition, in which different secondaries have different sets of allowed dressings. Could Stanley-type decompositions provide a useful organization of tree-level cohomology more generally?

\item
Can the secondaries be associated with semiclassical bulk backgrounds and the primaries with graviton excitations about them? The explicit nongraviton operators and their graviton dressings in the BMN sector provide a setting in which to test this interpretation~\cite{Choi:2023znd}. Can we identify graviton operators that act freely on these classes and determine whether finitely many secondaries suffice? Relating such a decomposition to bulk backgrounds would require understanding the states represented by the secondaries, beyond constructing an algebraic basis.

\item
How does the decomposition change with the gauge-group rank? Finite-$N$ trace identities can produce fortuitous classes that have no $Q$-closed lift before these identities are imposed~\cite{Chang:2024zqi}. Can a common family of primary dressings organize the cohomology at different ranks, with fortuitous classes appearing as additional rank-dependent secondaries? Our toy example realizes this possibility; it remains to determine whether gauge theories admit a similar description.

\item
Is there an appropriate quantum generalization of Hironaka decomposition? Loop corrections to $Q$ need not obey the Leibniz rule for the classical polynomial product, so this product need not descend to quantum cohomology~\cite{Budzik:2025zvu}. One must therefore first ask whether a corrected associative product compatible with the corrected supercharge exists. A weaker possibility is to identify mutually commuting dressing operations that are well-defined on quantum cohomology, without requiring a product between arbitrary classes. Can finitely many classes generate the quantum cohomology uniquely under these operations? This would retain the organizing role of Hironaka decomposition, although quantum corrections could change both the allowed dressings and the generating classes.
\end{itemize}

\acknowledgments

We thank Chi-Ming Chang, Jaehyeok Choi, Antal Jevicki, Robert de Mello Koch, Seok Kim, Eunwoo Lee and   Shiraz Minwalla for useful comments and discussions. We thank the Lodha Mathematical Sciences Institute for its hospitality during the meeting “Discussions on Quantum Spacetime,” where part of this work was carried out. We acknowledge the support of the Department of Atomic Energy, Government of India, under Project Identification No. RTI 4012. This work is partially supported by the Infosys Endowment for the Study of the Quantum Structure of Spacetime. The authors acknowledge their debt to the people of India for their steady support to the study of the basic sciences.

\paragraph{Statement on usage of AI} We used Claude Opus 5.5 (Anthropic) and GPT 6 Astra (OpenAI) in the preparation of this paper. \textbf{Code.} LLMs were used to write {\tt Mathematica} and {\tt Macaulay2} scripts. \textbf{Language.} LLMs were used to correct typos and ensure correct grammar. \textbf{Research Content.} The following ideas were developed in consultation with the AI (ChatGPT 6 Astra): 1. Counting the number of secondaries in diagonal models (eq. \ref{eq:sec_diagonal}) and in ABJM scalar ring (eq. \ref{eq:ABJMsc_count}) 2. Identifying primaries for the bosonic diagonal image of the SU(2) BMN graviton ring (eq. \ref{eq:BMN_grav}), N = 2 SU(2) SQCD with Nf = 4 (eq. \ref{eq:four-primaries}), the Klebanov–Witten scalar ring (eq. \ref{eq:kleb_prim}), and the restricted bi-fundamental complex motivated by ABJM theory (eq. \ref{eq:abjm_fer}). The research question and the main results are the authors’ own. The authors take full responsibility for the content of the paper.

\appendix
\section{Decomposition of the diagonal operator ring}
\label{app:diagonal}

For a single bosonic species, we first separate the symmetric-polynomial dressings from the remaining polynomial dependence on its $N$ components. Let $e_r$ be the elementary symmetric polynomial of degree $r$:
\begin{equation}
    e_r=\sum_{1\leq i_1<\cdots<i_r\leq N}
    x_{i_1}\cdots x_{i_r},\qquad r=1,\ldots,N.
\end{equation}
These generate the same polynomial algebra as the first $N$ power sums $\sum_i x_i^r$, so this change of primaries does not change the secondary count.

Consider the quotient obtained by imposing $e_1=...=e_N=0$,
\begin{equation}
    C_N=\mathbb C[x_1,\ldots,x_N]/(e_1,\ldots,e_N).
\end{equation}
The quotient identifies two polynomials if their difference is a sum of multiples of the $e_r$; it therefore describes the polynomial dependence that remains after the symmetric dressings have been removed. This space is called the coinvariant algebra. Its generating function is
\begin{equation}
    H_{C_N}(t)=\prod_{r=1}^{N}(1+t+\cdots+t^{r-1}),
\end{equation}
and hence it has dimension $N!$.

A standard result for symmetric polynomials~\cite{Chevalley:1955,Stanley:1979} states that a homogeneous basis of $C_N$ can be lifted to polynomials so that every polynomial in the $x_i$ has a unique expansion in these lifts, with coefficients that are  polynomial in the $e_r$:
\begin{equation}
    \mathbb C[x_1,\ldots,x_N]
    \simeq\mathbb C[e_1,\ldots,e_N]\otimes C_N.
\end{equation}
The basis can be organized by irreducible representation of $S_N$.

For example, take \(N=2\), with symmetric polynomials
\[
e_1=x_1+x_2,\qquad e_2=x_1x_2.
\]
After setting \(e_1=e_2=0\), the quotient has basis \(1,\delta\), where \(\delta=x_1-x_2\). Choosing these same polynomials before taking the quotient gives the ``lifted basis''. Under the permutation \(x_1\leftrightarrow x_2\),
\[
1\mapsto1,\qquad \delta\mapsto-\delta
\]
Thus 1 spans the trivial representation of $S_2$, while $\delta$ spans the sign representation. Every polynomial has a unique expansion
\[
F(x_1,x_2)=A(e_1,e_2)+\delta\,B(e_1,e_2),
\]
with polynomial coefficients \(A,B\). For example,
\[
x_1^2=\frac{e_1^2}{2}-e_2+\frac{e_1}{2}\delta.
\]
Products reduce to the same basis because
\[
\delta^2=(x_1-x_2)^2=e_1^2-4e_2.
\]
Thus multiplying two basis elements again gives a combination of \(1,\delta\), with symmetric-polynomial coefficients. Setting the primaries to zero recovers the quotient relation \(\delta^2=0\).

For one fermionic species, write
\begin{equation}
    \eta=\sum_{i=1}^{N}\theta_i,\qquad
    \xi_i=\theta_i-\frac{\eta}{N},\qquad
    \sum_{i=1}^{N}\xi_i=0.
\end{equation}
The sum $\eta$ is invariant under permutations, while the $\xi_i$ describe the remaining $N-1$ independent fermionic modes, which are permuted among themselves. Let $E_N$ denote the space of their products, including the identity. Since each independent fermionic mode can appear at most once, $E_N$ has dimension $2^{N-1}$. Every fermionic polynomial has a unique expansion in products from $E_N$, with coefficients of the form $a+b\eta$.

Now take $d$ bosonic species and $f$ fermionic species, and apply the preceding separation to each species. Before imposing permutation invariance, the remaining space is
\begin{equation}
    C_N^{\otimes d}\otimes E_N^{\otimes f},
\end{equation}
where the tensor products allow independent choices from each species. Let $S_N$ denote the group of permutations of the $N$ labels. The secondary space is its invariant part,
\begin{equation}
    K_{N;d,f}=\left[C_N^{\otimes d}\otimes E_N^{\otimes f}\right]^{S_N}.
\end{equation}
All the extracted dressings are already invariant, so the permutation condition acts only on this remaining space.

Choose a homogeneous basis $s_\alpha$ of $K_{N;d,f}$, and write $e_r^{(a)}$ for the elementary symmetric primaries of bosonic species $a$. Every invariant operator then has the unique expansion
\begin{equation}
    \mathcal O=\sum_\alpha
    \sum_{\epsilon_1,\ldots,\epsilon_f=0}^{1}
    F_{\alpha,\boldsymbol\epsilon}\!\left(e_r^{(a)}\right)
    \eta_1^{\epsilon_1}\cdots\eta_f^{\epsilon_f}s_\alpha,
\end{equation}
with polynomial coefficients $F_{\alpha,\boldsymbol\epsilon}$ and the fermionic sums written in the displayed order. The space $K_{N;d,f}$ is finite-dimensional, so this proves the existence of the required finite secondary basis.

To count the invariant states in $K_{N;d,f}$, we average the permutation operators over $S_N$. This average projects onto invariant states, so its trace is the dimension of the secondary space. The useful property of $C_N$ is that, when the polynomial grading is ignored, it carries the regular representation of $S_N$~\cite{Chevalley:1955,Stanley:1979}. This means that a basis can be labelled by the $N!$ permutations, with the group acting by left multiplication on those labels. The identity then fixes every basis element, while any other permutation fixes none, and consequently
\begin{equation}
    \operatorname{Tr}_{C_N}\sigma=
    \begin{cases}N!,&\sigma=1,\\0,&\sigma\neq1.\end{cases}
\end{equation}

For $d\geq1$, this makes every nonidentity term vanish in the average, and the identity contributes the full dimension of each factor, giving
\begin{equation}\label{eq:sec_diagonal}
\begin{aligned}
    \dim K_{N;d,f}
    &=\frac1{N!}\sum_{\sigma\in S_N}
    \bigl(\operatorname{Tr}_{C_N}\sigma\bigr)^d
    \bigl(\operatorname{Tr}_{E_N}\sigma\bigr)^f\\
    &=\frac{(N!)^d\,2^{f(N-1)}}{N!}
    =(N!)^{d-1}2^{f(N-1)}.
\end{aligned}
\end{equation}
This gives the secondary count directly from the operator decomposition.

We can retain the fermion number in the same calculation. If $w$ counts the total number of relative fermions, the identity contributes $(1+w)^{f(N-1)}$ from the fermionic factors, so the polynomial that counts secondaries by fermion number, summed over bosonic degrees, is
\begin{equation}
    P_{N;d,f}(1,w)=(N!)^{d-1}(1+w)^{f(N-1)}.
\end{equation}
In particular, the number of secondaries containing $F$ relative fermions is
\begin{equation}
    r_F=(N!)^{d-1}\binom{f(N-1)}{F},
    \qquad 0\leq F\leq f(N-1).
\end{equation}
This reproduces both the total counts and the fermion distributions used in Section~\ref{sec:commuting}.

Let $u_a$, $a=1,\ldots,d$, count each bosonic species and $w_b$, $b=1,\ldots,f$, count each fermionic species. The generating function then follows by averaging over permutations of the diagonal labels. For a permutation $\sigma$, let $c$ run over its disjoint cycles and let $\ell_c$ be the length of cycle $c$. Then
\begin{equation}
    H_{N;d,f}(\boldsymbol u,\boldsymbol w)
    =\frac1{N!}\sum_{\sigma\in S_N}\prod_c
    \frac{\prod_{b=1}^{f}\bigl[1-(-w_b)^{\ell_c}\bigr]}
    {\prod_{a=1}^{d}(1-u_a^{\ell_c})}.
\end{equation}
Here $\boldsymbol u$ and $\boldsymbol w$ denote the lists of species weights.

The cycle factors have a direct explanation. A bosonic monomial contributes to the trace of a cycle only when every variable in that cycle has the same exponent, and summing these exponents gives $1/(1-u_a^{\ell_c})$. For one fermionic species, the cycle fixes only the empty product and the product containing all its $\ell_c$ fermions; reordering the latter contributes the sign $(-1)^{\ell_c-1}$, giving $1-(-w_b)^{\ell_c}$. Averaging these traces projects onto invariant operators, each counted with positive multiplicity.

Removing the primary dressings gives the secondary polynomial,
\begin{equation}
    P_{N;d,f}(\boldsymbol u,\boldsymbol w)
    =H_{N;d,f}(\boldsymbol u,\boldsymbol w)\,
    \frac{\displaystyle\prod_{a=1}^{d}\prod_{r=1}^{N}(1-u_a^r)}
    {\displaystyle\prod_{b=1}^{f}(1+w_b)}.
\end{equation}
The decomposition proved above ensures that this expression is a polynomial with nonnegative integer coefficients, which give the number of secondaries at each set of species degrees. Taking $(d,f)=(2,1)$ and $(3,2)$ gives the $1/4$-BPS and $1/8$-BPS numerators, respectively.

\subsection{Relative oscillator states and bosonic numerators}
\label{app:oscillators}

Consider two identical bosons in a $d$-dimensional isotropic harmonic oscillator. Their positions $r_1,r_2$ separate into a center-of-mass coordinate $R=(r_1+r_2)/2$ and a relative coordinate $r=r_1-r_2$. Both oscillators have the original frequency. Let $t$ count excitation quanta, with the vacuum energy removed. The center of mass contributes $(1-t)^{-d}$, while exchange symmetry requires the relative wavefunction to obey $\Psi(-r)=\Psi(r)$. We use the usual oscillator wavefunctions, regular at $r=0$.

In one dimension, the even wavefunctions have excitation numbers $2n$, $n\geq0$, giving a relative partition function $(1-t^2)^{-1}$. In two dimensions, exchange shifts the polar angle by $\pi$, so the angular momentum $\ell$ must be even. With radial quantum number $n\geq0$, the excitation number is $2n+|\ell|$, and hence
\begin{equation}
 H_{\rm rel}^{(2)}(t)=\sum_{n\geq0}\sum_{\ell\in2\mathbb Z}
 t^{2n+|\ell|}=\frac{1+t^2}{(1-t^2)^2}.
\end{equation}
In three dimensions, an angular momentum $\ell$ has parity $(-1)^\ell$ and degeneracy $2\ell+1$. Keeping $\ell=2j$, with $j\geq0$, gives
\begin{equation}
 H_{\rm rel}^{(3)}(t)=\sum_{n,j\geq0}(4j+1)t^{2n+2j}
 =\frac{1+3t^2}{(1-t^2)^3}.
\end{equation}
Multiplication by the center-of-mass factor reproduces the two-particle bosonic generating function. The numerators count the relative states left after separating the primary dressings, in agreement with the bosonic diagonal model.

For two bosonic species at higher rank, let $P_N^{\rm bos}(t)$ be the unrefined numerator over $\prod_{r=1}^N(1-t^r)^2$. The permutation average derived above gives
\begin{align}
 P_3^{\rm bos}(t)&=1+t^2+2t^3+t^4+t^6,\notag\\
 P_4^{\rm bos}(t)&=1+t^2+2t^3+4t^4+2t^5+4t^6+2t^7
                  +4t^8+2t^9+t^{10}+t^{12}.
\end{align}
Their coefficient sums are $3!$ and $4!$, as required by the general rank formula. The extra terms measure independent relative excitations that cannot be generated by the separate spectral primaries alone.

\section{\texorpdfstring{A $1/4$-BPS basis for $U(3)$}{A 1/4-BPS basis for U(3)}}
\label{app:u3}

For the $U(3)$ $1/4$-BPS model, write the diagonal components of $X,Y,\Psi$ as $x_i,y_i,\theta_i$, with $i=1,2,3$. We choose the six bosonic primaries $\operatorname{Tr}X^r,\operatorname{Tr}Y^r$, $r=1,2,3$, and the fermionic dressing $\eta=\operatorname{Tr}\Psi$. After removing these dressings, the secondary space contains six operators with no fermions, twelve with one fermion and six with two fermions.

Let $u,v,w$ count $X,Y,\Psi$, respectively, and write the secondary polynomial as
\begin{equation}
    P_3(u,v,w)=P_3^{(0)}(u,v)
    +wP_3^{(1)}(u,v)+w^2P_3^{(2)}(u,v).
\end{equation}
The superscript records the fermion number, and the zero-fermion part is
\begin{equation}
    P_3^{(0)}(u,v)=1+uv+uv^2+u^2v+u^2v^2+u^3v^3.
\end{equation}
A corresponding choice of six scalar secondaries is
\begin{equation}
\begin{aligned}
\mathcal S^{(0)}=\{&
1,\ \operatorname{Tr}(XY),\
\operatorname{Tr}(XY^2),\
\operatorname{Tr}(X^2Y),\\
&[\operatorname{Tr}(XY)]^2,\
[\operatorname{Tr}(XY)]^3\}.
\end{aligned}
\end{equation}
Their weights reproduce the six terms in $P_3^{(0)}$. Compared with $U(2)$, the additional eigenvalue allows more independent mixed traces and products. We verify the independence of this list below, together with the fermionic lists introduced next.

To keep the lists short, define
\begin{equation}
T_{mn}=\operatorname{Tr}(X^mY^n),\qquad
F_{mn}=\operatorname{Tr}(X^mY^n\Psi),
\end{equation}
where $m,n$ are nonnegative integers. We choose the twelve one-fermion secondaries as
\begin{equation}
\begin{aligned}
\mathcal S^{(1)}=\{&
F_{10},F_{01},F_{20},F_{02},F_{11},\\
&T_{11}F_{10},T_{11}F_{01},
T_{11}F_{20},T_{11}F_{02},\\
&T_{12}F_{10}+T_{21}F_{01},
T_{12}F_{11},T_{21}F_{11}\}.
\end{aligned}
\end{equation}
The six two-fermion secondaries are
\begin{equation}
\begin{aligned}
\mathcal S^{(2)}=\{&
F_{10}F_{20},F_{01}F_{02},F_{10}F_{01},\\
&F_{10}F_{11},F_{01}F_{11},F_{20}F_{02}\}.
\end{aligned}
\end{equation}
The $F_{mn}$ anticommute, and their products are ordered as written.

Counting the powers of $X$ and $Y$ in these lists gives
\begin{equation}
\begin{aligned}
P_3^{(1)}(u,v)={}&
u+v+u^2+v^2+uv
+u^2v+uv^2\\
&+u^3v+uv^3+u^2v^2
+u^2v^3+u^3v^2,\\
P_3^{(2)}(u,v)={}&
u^3+v^3+uv+u^2v+uv^2+u^2v^2.
\end{aligned}
\end{equation}
These agree with the one- and two-fermion terms of the generating function and, together with the scalar list, give the required $6+12+6=24$ secondaries.

To check independence, we remove the dressings by setting all six bosonic primaries and $\eta$ to zero. The linear relations give
\begin{equation}
x_3=-x_1-x_2,\qquad
y_3=-y_1-y_2,\qquad
\theta_3=-\theta_1-\theta_2.
\end{equation}
The remaining scalar relations can be written as
\begin{equation}
\begin{aligned}
x_1^2+x_1x_2+x_2^2&=0,& x_2^3&=0,\\
y_1^2+y_1y_2+y_2^2&=0,& y_2^3&=0.
\end{aligned}
\end{equation}
Using these relations, we expand each proposed secondary in monomials of the remaining variables. Computed with exact rational arithmetic, the coefficient matrices have ranks $6,12,6$ in the zero, one and two fermion sectors, so each list is independent. Since their dimensions agree with the generating function, they also span the space left after removing the dressings.

Restoring the primaries now gives a Hironaka decomposition. Indeed, any operator can first be reduced to a linear combination of the listed secondaries after setting the dressings to zero. The difference contains at least one dressing, whose coefficient has lower degree, and repeating the reduction expresses every operator in terms of dressed secondaries. Since their generating function equals that of the full operator space at every degree, this expansion is unique.

\subsection{A secondary-product reduction}

To reduce products in this basis, we need trace identities that retain the primary dependence. Define
\begin{equation}
\begin{gathered}
 S=T_{11},\quad T=T_{21},\quad U=T_{12},\\
 A=F_{10},\quad B=F_{01},\quad C=F_{20},\quad D=F_{11},\quad E=F_{02},\\
 p_r=\operatorname{Tr}X^r,\qquad q_r=\operatorname{Tr}Y^r,\qquad
 \eta=\operatorname{Tr}\Psi.
\end{gathered}
\end{equation}
With the fermionic dressing written on the left of bosonic traces, one finds the polynomial identity
\begin{equation}
\begin{aligned}
 TA={}&-SC+\frac43p_1SA+\left(p_2-\frac13p_1^2\right)D
       +\left(p_3-\frac43p_1p_2+\frac13p_1^3\right)B\\
 &+\frac13p_1q_1C-\frac13p_1^2q_1A
   +\frac13p_1\eta T-\frac13p_1^2\eta S
   -\frac13p_3q_1\eta+\frac13p_1p_2q_1\eta.
\end{aligned}
\end{equation}
Exchanging $X$ and $Y$ gives the corresponding reduction of $UB$: $p_r\leftrightarrow q_r$, $T\leftrightarrow U$, $A\leftrightarrow B$, $C\leftrightarrow E$, while $S,D,\eta$ remain unchanged. Both identities hold before taking the primary quotient. In that quotient they reduce to $TA=-SC$ and $UB=-SE$, explaining why $TA$ and $UB$ need not be added to the listed one-fermion basis.  Further secondary-product reduction relations are provided in the ancillary Mathematica notebook \texttt{\detokenize{1-4BPS_U3_secondary_relations_revised_s1_s24.nb}}.

\section{\texorpdfstring{A $1/8$-BPS basis for $U(2)$}{A 1/8-BPS basis for U(2)}}
\label{app:u2-eighth}

For the diagonal $U(2)$ $1/8$-BPS model, write the components of $X,Y,Z$ as $x_i,y_i,z_i$ and those of $\Psi_A$ as $\theta_{iA}$, with $i=1,2$ and $A=1,2$. We choose the six bosonic primaries $\operatorname{Tr}X^k,\operatorname{Tr}Y^k,\operatorname{Tr}Z^k$, $k=1,2$, and the fermionic dressings $\eta_A=\operatorname{Tr}\Psi_A$. After removing these dressings, the secondary space contains four operators with no fermions, eight with one fermion and four with two fermions.

Let $u_1,u_2,u_3$ count $X,Y,Z$, respectively, and let $w$ count the total number of fermions of both species. Write the secondary polynomial as
\begin{equation}
\begin{aligned}
P_2^{1/8}(u_1,u_2,u_3,w)
={}&P_2^{(0)}(u_1,u_2,u_3)\\
&+(w_1+w_2)P_2^{(1)}(u_1,u_2,u_3)
+w_1w_2P_2^{(2)}(u_1,u_2,u_3).
\end{aligned}
\end{equation}
The superscript records the total fermion number, and the zero-fermion part is
\begin{equation}
P_2^{(0)}(u_1,u_2,u_3)=1+u_1u_2+u_2u_3+u_3u_1.
\end{equation}
To keep the lists short, define
\begin{equation}
\begin{aligned}
T_{mnp}&=\operatorname{Tr}(X^mY^nZ^p),\\
F^A_{mnp}&=\operatorname{Tr}(X^mY^nZ^p\Psi_A),\\
G_{mnp}&=\operatorname{Tr}(X^mY^nZ^p\Psi_1\Psi_2),
\end{aligned}
\end{equation}
where $m,n,p$ are nonnegative integers. The fermions in $G_{mnp}$ are ordered as written.

We choose the scalar secondaries as
\begin{equation}
\mathcal S^{(0)}=\{1,\ T_{110},\ T_{011},\ T_{101}\},
\end{equation}
the one-fermion secondaries as
\begin{equation}
\mathcal S^{(1)}
=
\{F^1_{100},\ F^1_{010},\ F^1_{001},\ F^1_{111},\ F^2_{100},\ F^2_{010},\ F^2_{001},\ F^2_{111}\},
\end{equation}
and the two-fermion secondaries as
\begin{equation}
\mathcal S^{(2)}=\{G_{000},\ G_{110},\ G_{011},\ G_{101}\}.
\end{equation}
Counting the powers of $X,Y,Z$ in these lists gives
\begin{equation}
\begin{aligned}
P_2^{(1)}(u_1,u_2,u_3)&=(u_1+u_2+u_3+u_1u_2u_3),\\
P_2^{(2)}(u_1,u_2,u_3)&=1+u_1u_2+u_2u_3+u_3u_1.
\end{aligned}
\end{equation}
The two copies of $P_2^{(1)}$ are due to the two fermion species. These agree with the one- and two-fermion terms of the generating function and, together with the scalar list, give the required $4+8+4=16$ secondaries.

To check independence, set all six bosonic primaries and both $\eta_A$ to zero. The linear relations give
\begin{equation}
x_2=-x_1,\qquad y_2=-y_1,\qquad z_2=-z_1,\qquad
\theta_{2A}=-\theta_{1A},
\end{equation}
while the remaining scalar relations are
\begin{equation}
x_1^2=y_1^2=z_1^2=0.
\end{equation}
Each listed trace reduces to twice the monomial obtained from its first diagonal entry. Together with the identity, these are distinct nonzero monomials in the quotient, so the three lists are independent. Since their dimensions agree with the generating function, they also span the space left after removing the dressings.

Restoring the primaries gives a Hironaka decomposition. Repeated reduction modulo the dressings expresses every operator in the listed secondaries: each remainder contains a dressing whose coefficient has lower total degree. The coefficients are polynomial in the bosonic primaries and at most linear in each $\eta_A$. Since the dressed generating function equals that of the full operator space, this expansion is unique.

\subsection{A secondary-product reduction}

To retain the primary dependence, write
\begin{equation}
p=\operatorname{Tr}X,\qquad q=\operatorname{Tr}Y,\qquad
r=\operatorname{Tr}Z.
\end{equation}
With the fermionic dressing written on the left of the secondaries, one finds
\begin{equation}
\begin{aligned}
T_{110}F^A_{001}={}&2F^A_{111}
-\frac12q\eta_A T_{101}
-\frac12p\eta_A T_{011}\\
&-\frac12qrF^A_{100}
-\frac12prF^A_{010}
+pqr\eta_A,
\qquad A=1,2.
\end{aligned}
\end{equation}
Permuting $X,Y,Z$ gives the corresponding reductions for the other scalar pairs. These identities hold before taking the primary quotient. In that quotient they reduce to $T_{110}F^A_{001}=2F^A_{111}$ and its permutations, explaining why these products need not be added to the one-fermion basis. The complete secondary-product relations are provided in the ancillary Mathematica notebook \texttt{\detokenize{1-8BPS_U2_secondary_relations.nb}}

\section{A basis for the bosonic BMN diagonal image}
\label{app:bmn}

We now give the primaries and secondaries for the bosonic diagonal ring discussed in the main text. Write the traceless diagonal matrices as
\begin{equation}
X=\operatorname{diag}(x,-x),\qquad
Y=\operatorname{diag}(y,-y),
\end{equation}
and similarly for $Z$ and $F$, with entries $z$ and $f$. The ring is generated by
\begin{equation}
x^2,\ y^2,\ z^2,\ xy,\ yz,\ zx,\ xf,\ yf,\ zf.
\end{equation}
Every monomial in these generators has at least as many scalar factors as factors of $f$, and an even total number of factors. Conversely, any monomial satisfying these conditions can be built from the generators: pair each $f$ with a scalar, then pair the remaining scalars. Thus a monomial basis is
\begin{equation}
x^a y^b z^c f^d,\qquad
a,b,c,d\geq0,\quad
a+b+c\geq d,\quad
a+b+c+d\ \text{even}.
\end{equation}
These exponent vectors are precisely the points of the even-total-degree lattice inside the cone $a+b+c\geq d$, $a,b,c,d\geq0$. The cone is defined by finitely many homogeneous linear inequalities with integer coefficients and contains no line. For a cone of this type, the ring containing every lattice monomial in the cone is Cohen--Macaulay and therefore admits a Hironaka decomposition~\cite{Hochster:1972}. The remaining task is to exhibit one. The numerator found below is palindromic, the symmetry discussed in Section~\ref{sec:hironaka}.

We assign degree two to $x,y,z$ and degree six to $f$, as in the main text. The six scalar generators then have degree four, while $xf,yf,zf$ have degree eight. We will exhibit four primaries of degrees $4,8,8,8$ and 21 secondaries, and check that their products reproduce this monomial basis at every degree.

We choose
\begin{equation}\label{eq:BMN_grav}
\begin{aligned}
m_1={}&xy-y^2,\\
m_2={}&y^2z^2-y^3z+y^4-xf
-xyz^2-xy^2z+xy^3\\
&+x^2yz+x^2y^2+x^3z-x^3y,\\
m_3={}&-zf+z^4-yf+yz^3+y^4-xf+xy^3\\
&+x^2z^2-x^2yz+x^2y^2+x^4,\\
m_4={}&-z^4+yf-yz^3-y^2z^2+y^3z+xf+xz^3\\
&+xyz^2-xy^2z+xy^3-x^2y^2+x^3z+x^4.
\end{aligned}
\end{equation}
Each term belongs to the diagonal ring. The first primary has degree four, while the other three combine scalar quartics with the degree-eight terms $xf,yf,zf$. This choice allows all 21 secondaries to be individual monomials; we give their list below and then verify that every element of the ring has a unique expansion in these secondaries with polynomial coefficients in $m_1,m_2,m_3,m_4$.

Let $\mathcal S_D$ denote the secondaries of degree $D$. We choose
\begin{equation}
\begin{aligned}
\mathcal S_0={}&\{1\},\\
\mathcal S_4={}&\{z^2,yz,y^2,xz,x^2\},\\
\mathcal S_8={}&\{zf,z^4,yf,yz^3,y^2z^2,y^3z,y^4,xf,xz^3\},\\
\mathcal S_{12}={}&\{z^3f,z^6,yz^2f,yz^5,y^2z^4\},\\
\mathcal S_{16}={}&\{z^2f^2\}.
\end{aligned}
\end{equation}
The generator--secondary reduction identities and secondary multiplication relations are provided in the ancillary Mathematica notebook \texttt{\detokenize{SU2_BMN_0fermion_graviton.nb}}.

Their numbers at each degree give the numerator
\begin{equation}
1+5t^4+9t^8+5t^{12}+t^{16}.
\end{equation}
Allowing arbitrary powers of the four primaries therefore gives the counting function
\begin{equation}
\frac{1+5t^4+9t^8+5t^{12}+t^{16}}
{(1-t^4)(1-t^8)^3},
\end{equation}
which agrees with the Hilbert series of the diagonal ring. To establish the decomposition, we must also check that the dressed secondaries span the ring; once this is shown, equality of the counting functions ensures that the expansion is unique.

At each degree $D$, we expand all products
\begin{equation}
s\,m_1^{n_1}m_2^{n_2}m_3^{n_3}m_4^{n_4},
\qquad n_i\geq0,
\end{equation}
of that degree in the monomial basis, with $s$ drawn from the secondary lists above. The coefficients form a square integer matrix. Through degree 24, its sizes and ranks are
\begin{equation}
\begin{array}{c|rrrrrrr}
D&0&4&8&12&16&20&24\\ \hline
\text{size}&1&6&18&38&72&117&184\\
\text{rank}&1&6&18&38&72&117&184
\end{array},
\end{equation}
where the ranks were computed modulo the prime $1000003$. Full rank modulo a prime implies that the integer determinant is nonzero, so these products span the ring at each listed degree over $\mathbb C$.

This check also establishes spanning at higher degrees. Every generator has degree at most eight and every secondary has degree at most sixteen, so their product has degree at most 24 and can be expressed in dressed secondaries by the calculation above. Multiplying any dressed secondary by a generator consequently gives another linear combination of dressed secondaries, and since the list includes $1$, repeated multiplication reaches every monomial in the ring.

Finally, at every degree the number of proposed dressed secondaries equals the dimension given by the Hilbert series, so spanning implies independence. Every element therefore has a unique expansion
\begin{equation}
\mathcal O=\sum_s s\,A_s(m_1,m_2,m_3,m_4),
\end{equation}
where each $A_s$ is a polynomial. This proves the Hironaka decomposition of the bosonic diagonal ring.

\section{Decomposition of the ABJM scalar-branch ring}
\label{app:abjm}

For $N$ branes, the scalar-branch ring consists of polynomials in the coordinates $z_i^{(a)}$ that are invariant under the discrete phases and brane permutations. We use the primaries
\begin{equation}
p_r^{(a)}=\sum_{i=1}^N\bigl(z_i^{(a)}\bigr)^{kr},
\qquad a=1,\ldots,4,\quad r=1,\ldots,N.
\end{equation}
To see how these dressings control high powers of the coordinates, introduce
\begin{equation}
q_a(T)=\prod_{i=1}^N
\left[T^k-\bigl(z_i^{(a)}\bigr)^k\right],
\end{equation}
where $T$ is an auxiliary variable. The coefficients are symmetric polynomials in the $k$th powers of the coordinates, so they can be expressed in the chosen primaries. Setting $T=z_j^{(a)}$ gives a relation that expresses $\bigl(z_j^{(a)}\bigr)^{kN}$ in lower powers, with primary-polynomial coefficients.

Applying these relations repeatedly reduces any polynomial to a sum of monomials whose individual coordinate exponents are less than $kN$, multiplied by polynomials in the primaries. If the polynomial is invariant, we can then average this expression over the discrete phases and brane permutations: the primaries remain unchanged, while the remaining monomials become invariant combinations. Finitely many such combinations therefore suffice to generate the scalar-branch ring over the chosen dressings. We next show that they can be chosen to give a unique expansion.

First separate the remainder of each coordinate exponent upon division by $k$. Writing $u_i^{(a)}=(z_i^{(a)})^k$, every polynomial has a unique expression
\begin{equation}
\sum_{\boldsymbol\rho}
\left[\prod_{a=1}^4\prod_{i=1}^N
(z_i^{(a)})^{\rho_i^{(a)}}\right]
B_{\boldsymbol\rho}\bigl(u_i^{(a)}\bigr),
\qquad 0\leq\rho_i^{(a)}<k.
\end{equation}
Here $\boldsymbol\rho$ records all the remainders, and each $B_{\boldsymbol\rho}$ is a polynomial in the $k$th powers. For each species $a$, we can now apply the symmetric-polynomial decomposition of Appendix~\ref{app:diagonal} to the $N$ variables $u_i^{(a)}$, which extracts the symmetric dressings and leaves $N!$ basis elements. Since the first $N$ power sums $p_r^{(a)}$ generate the same symmetric-polynomial ring as the elementary symmetric polynomials, they give the same unique expansion.

Combining the remainder monomials with these basis elements leaves a finite-dimensional space of polynomials, which we denote by $V$. The basis in Appendix~\ref{app:diagonal} can be chosen so that permutations act within its span, and consequently brane permutations preserve $V$. The discrete phases also preserve it, since they multiply each remainder monomial by a phase and leave every $u_i^{(a)}$ unchanged. All primary dressings are invariant under both operations.

It therefore suffices to select the invariant combinations in $V$. Choose a basis $s_\alpha$ of these combinations, with each element of definite degree. Every invariant polynomial then has the unique expansion
\begin{equation}
\mathcal O=\sum_\alpha
s_\alpha\,A_\alpha\!\left(p_r^{(a)}\right),
\end{equation}
where the $A_\alpha$ are polynomials. This proves the Hironaka decomposition over the primaries used in the main text.

\section{Class-S scalar bases and counting formulas}
\label{app:classs}
\subsection{The four-puncture ring}

We use the 28 antisymmetric moment maps $m_{AB}$ defined in Section~\ref{sec:classs}. Assemble them into the antisymmetric matrix $\mu$, with $\mu_{AB}=m_{AB}$ for $A<B$. The ideal $I$ is generated by the 70 Pl\"ucker quadrics and the 36 independent entries of $\mu^2$:
\begin{equation}
\begin{aligned}
 m_{AB}m_{CD}-m_{AC}m_{BD}+m_{AD}m_{BC}&=0,
                  &&A<B<C<D,\\
 \sum_{C=1}^8\mu_{AC}\mu_{CB}&=0,
                  &&A\leq B.
\end{aligned}
\end{equation}
The Higgs-branch ring is thus $R_4=\mathbb C[m_{AB}]/I$. This minimal nilpotent orbit closure has a Cohen--Macaulay and Gorenstein coordinate ring~\cite{Beauville_2000}. It therefore admits a Hironaka decomposition and has a palindromic numerator, as explained in Section~\ref{sec:hironaka}. We use this property only after checking that the particular proposed primaries have finite quotient.

Choose
\begin{equation}
\label{eq:four-primaries}
\begin{aligned}
 p_1&=m_{12}+m_{36}+m_{45},&p_2&=m_{13}+m_{34}-m_{35},\\
 p_3&=m_{14}+m_{34}-m_{68},&p_4&=m_{15}+m_{28}+m_{48},\\
 p_5&=m_{16}-m_{27}+m_{37},&p_6&=m_{17}+m_{26}-m_{27},\\
 p_7&=m_{18}+m_{58}+m_{78},&p_8&=m_{23}-m_{46}-m_{68},\\
 p_9&=m_{24}+m_{35}+m_{67},&p_{10}&=m_{25}-m_{58}+m_{78}.
\end{aligned}
\end{equation}
We describe the entire secondary list by six sets $\mathcal S_r$, where $r$ is field degree, so that a moment map has degree two. The first two sets are $\mathcal S_0=\{1\}$ and
\begin{equation}
\begin{aligned}
 \mathcal S_2=\{&m_{26},m_{27},m_{28},m_{34},m_{35},m_{36},m_{37},m_{38},m_{45},\\
               &m_{46},m_{47},m_{48},m_{56},m_{57},m_{58},m_{67},m_{68},m_{78}\}.
\end{aligned}
\end{equation}
The 65 elements of $\mathcal S_4$ are given in Table~\ref{tab:four-secondaries}: in each row, multiply the left-hand entry by each entry on the right. The higher-degree sets are
\begin{equation}
\label{eq:four-secondary-sets}
 \mathcal S_6=m_{78}\mathcal S_4,\qquad
 \mathcal S_8=m_{78}^3\mathcal S_2,\qquad
 \mathcal S_{10}=\{m_{78}^5\}.
\end{equation}
These prescriptions give all $1+18+65+65+18+1=168$ operators without repeating the same products at successive degrees.

\begin{table}[t]
\centering
\begin{tabular}{c|l}
First factor & Allowed second factors\\\hline
$m_{26}$ & $m_{67},m_{78}$\\
$m_{27}$ & $m_{67},m_{68},m_{78}$\\
$m_{28}$ & $m_{58},m_{68},m_{78}$\\
$m_{34},m_{35},m_{36}$ & $m_{58},m_{67},m_{68},m_{78}$\\
$m_{37}$ & $m_{57},m_{58},m_{67},m_{68},m_{78}$\\
$m_{38}$ & $m_{58},m_{68},m_{78}$\\
$m_{45},m_{46},m_{47}$ & $m_{57},m_{58},m_{67},m_{68},m_{78}$\\
$m_{48}$ & $m_{58},m_{68},m_{78}$\\
$m_{56},m_{57}$ & $m_{57},m_{58},m_{67},m_{68},m_{78}$\\
$m_{58}$ & $m_{58},m_{68},m_{78}$\\
$m_{67}$ & $m_{67},m_{68},m_{78}$\\
$m_{68}$ & $m_{68},m_{78}$\\
$m_{78}$ & $m_{78}$
\end{tabular}
\caption{The 65 degree-four secondaries of the four-puncture ring.
When several first factors share a row, each is multiplied by every
listed second factor.}
\label{tab:four-secondaries}
\end{table}

To verify the basis, set the ten primaries to zero and eliminate the seven entries $m_{1B}$, $2\leq B\leq8$, and $m_{23},m_{24},m_{25}$ using \eqref{eq:four-primaries}. Denote the resulting quotient by $\overline R_4$. Gr\"obner reduction of all the quadrics over $\mathbb Q$ gives its Hilbert function, and the normal forms of the listed operators have the same ranks in each degree:
\begin{equation}
\begin{array}{c|rrrrrrr}
 \text{moment-map degree}&0&1&2&3&4&5&6\\\hline
 \dim(\overline R_4)&1&18&65&65&18&1&0\\
 \text{rank of listed operators}&1&18&65&65&18&1&0
\end{array}.
\end{equation}
The quotient is generated in moment-map degree one, so vanishing in degree six implies vanishing in every higher degree. Since $R_4$ has dimension ten, finiteness of this quotient shows that the ten primaries are algebraically independent and leave only finitely many operators when set to zero. Since $R_4$ admits a Hironaka decomposition over any such primaries (Section~\ref{sec:hironaka}), the quotient basis lifts to a Hironaka basis, which proves the decomposition in every degree. Secondary products begin to reduce, for example, through
\begin{equation}
 m_{35}m_{68}-m_{36}m_{58}+m_{38}m_{56}=0,\qquad
 m_{26}m_{78}-m_{27}m_{68}+m_{28}m_{67}=0.
\end{equation}

The primary-quotient calculation and secondary-basis checks are implemented in the ancillary Macaulay2 script \texttt{\detokenize{4punctures_secondaries.m2}}.

\subsection{More punctures}
\label{app:more-punctures}

For the untwisted $A_1$ sphere with $s$ full punctures, the Hall--Littlewood expression of~\cite{Gadde:2011uv} can be written as
\begin{equation}
\label{eq:hl-refined}
 H_s(t;\boldsymbol a)=\frac{(1+t^2)^{s-2}}{1-t^2}
 \sum_{\lambda=0}^\infty
 \frac{\displaystyle\prod_{I=1}^s
 \frac{P_\lambda(a_I\mid t)}{(1-t^2a_I^2)(1-t^2a_I^{-2})}}
 {P_\lambda(t\mid t)^{s-2}}.
\end{equation}
The $a_I$ are $SU(2)$ flavor fugacities and $t$ counts one hypermultiplet scalar. Here
\begin{equation}
 P_0(a\mid t)=\sqrt{1+t^2},\qquad
 P_\lambda(a\mid t)=\chi_\lambda(a)-t^2\chi_{\lambda-2}(a)
 \quad(\lambda\geq1),
\end{equation}
where $\chi_\lambda(a)=(a^{\lambda+1}-a^{-\lambda-1})/(a-a^{-1})$ is the $SU(2)$ character and $\chi_{-1}=0$. In particular $\chi_0=1$; the factor $\sqrt{1+t^2}$ belongs to $P_0$, not to the trivial character. Since a Hironaka basis need not preserve the individual flavor weights, we first remove the flavor refinement and ask whether the resulting series admits a positive finite numerator. Taking $a_I\to1$ gives
\begin{equation}
\label{eq:hl-unrefined}
 H_s(t)=\frac{(1+t^2)^{s-1}+\displaystyle\sum_{\lambda\geq1}
 t^{\lambda(s-2)}[(\lambda+1)-t^2(\lambda-1)]^s}
 {(1-t^2)^{2s+1}}.
\end{equation}
This is a rational function: after expanding the bracket, each term is a sum of $\lambda^j x^\lambda$ with $x=t^{s-2}$, obtained by applying $(x\partial_x)^j$ to $x/(1-x)$.

For $s=5,6,7$, define $N_s(t)$ by
\begin{equation}
 H_s(t)=\frac{N_s(t)}{(1-t^2)^{s+1}(1-t^{s-2})^{s+1}}.
\end{equation}
The numerators are palindromic, meaning that the coefficients at degrees $r$ and $D_s-r$ agree, where $D_s=\deg N_s$. It is economical to give them as
\begin{equation}
 N_s(t)=L_s(t)+c_st^{D_s/2}+t^{D_s}L_s(t^{-1}),
\end{equation}
with
\begin{align}
 L_5(t)&=1+9t^2+26t^3+41t^4+106t^5+195t^6+234t^7+306t^8,
 & (D_5,c_5)&=(18,372),\notag\\
 L_6(t)&=1+11t^2+118t^4+538t^6+1900t^8+4109t^{10}+6901t^{12},
 & (D_6,c_6)&=(28,7804).
\end{align}
For seven punctures the corresponding data are
\begin{equation}
\begin{aligned}
 L_7(t)={}&1+13t^2+85t^4+120t^5+377t^6+792t^7+1289t^8
            +2904t^9\\
         &+4844t^{10}+7736t^{11}+12391t^{12}+16568t^{13}
            +24151t^{14}\\
         &+32328t^{15}+38331t^{16}+46416t^{17}
            +51171t^{18}+55056t^{19},\\
 (D_7,c_7)={}&(40,59094).
\end{aligned}
\end{equation}
The sums $N_s(1)$ are $2208$, $34960$ and $648240$, respectively, and would count secondaries for primary algebras with the displayed degrees, if such parameter choices were established. The five-puncture presentation in particular has six degree-two and six degree-three denominator factors, and all numerator coefficients are nonnegative.

We evaluated \eqref{eq:hl-unrefined} as an exact rational function for $4\leq s\leq20$, and suitable denominator rewritings give positive palindromic numerators in every case checked. The compact degree pattern above is not universal, however: it fails to give a polynomial numerator already at $s=9$. Where needed, taking all $2s+2$ denominator degrees equal to the least common multiple $\operatorname{lcm}(2,s-2)$ gives the checked rewriting. These are counting identities, not constructions of primary operators. Higher-rank fixtures also supply examples, such as the $T_3$ fixture with $E_6$ flavor symmetry~\cite{Minahan:1996fg,Gadde:2011uv}; their operator bases require separate work. 

\subsection{The closed genus-two surface}
\label{app:genus-two}

The dumbbell frame has three $SU(2)$ gauge groups, associated with its left loop, joining edge and right loop. Write its matter tensors as $q_{ab i}$ and $\widetilde q_{cd i}$: the first two indices of each tensor belong to its loop group, and $i$ belongs to the joining group. All indices run from one to two. Define
\begin{equation}
 \alpha_i=\epsilon^{ab}q_{ab i},\qquad
 \beta_i=\epsilon^{cd}\widetilde q_{cd i},\qquad
 U_{ij}=\epsilon^{aa'}\epsilon^{bb'}q_{ab i}q_{a'b'j}.
\end{equation}
The matrix $U$ is symmetric, and a normalized set of generators is
\begin{equation}
 M=\frac12\epsilon^{ij}\alpha_i\beta_j,\qquad
 B_1=\epsilon^{ik}\epsilon^{jl}\alpha_iU_{kl}\alpha_j,\qquad
 B_2=\epsilon^{ik}\epsilon^{jl}\alpha_iU_{kl}\beta_j.
\end{equation}
The scalar F-term identities in this frame imply $\det U=B_1/2$ and $\epsilon^{ik}\epsilon^{jl}\beta_iU_{kl}\beta_j=-B_1$ \cite{Hanany_2011}. The two-dimensional determinant identity therefore gives $4M^2\det U+B_2^2+B_1^2=0$, which is \eqref{eq:genus-two-ring}. Our factor of $1/2$ in $M$ fixes the coefficient of its term in this relation.

Polynomial division in $B_2$ proves that the basis over $\mathbb C[M,B_1]$ is $\{1,B_2\}$. Equivalently, the Hilbert series is
\begin{equation}
 \frac{1-t^8}{(1-t^2)(1-t^4)^2}
 =\frac{1+t^4}{(1-t^2)(1-t^4)}.
\end{equation}
The independent Hilbert-series calculation in the Yin--Yang frame agrees~\cite{Hanany_2011}. This is a scalar check of the two descriptions, without assuming a corresponding identification of fermionic bases.

\section{Polynomial cohomology and gauged examples}
\label{app:cohomology}
\subsection{The Koszul differential}

Let $z_a$ be commuting scalar variables, $\psi_\alpha$ Grassmann-odd variables, and $f_\alpha(z)$ scalar polynomials. The complex with $Qz_a=0$ and $Q\psi_\alpha=f_\alpha$ has chain spaces
\begin{equation}
 C_j=\mathbb C[z_a]\otimes\Lambda^j\langle\psi_\alpha\rangle,
 \qquad Q=\sum_\alpha f_\alpha(z)\frac{\partial}{\partial\psi_\alpha}.
\end{equation}
Here $\Lambda^j\langle\psi_\alpha\rangle$ is the space of products of $j$ distinct fermions. The derivatives are left Grassmann derivatives, $Q^2=0$, and $Q$ lowers fermion number by one. The cohomology $\mathcal H_j=\ker(Q:C_j\to C_{j-1})/\operatorname{im}(Q:C_{j+1}\to C_j)$ is the Koszul homology of the scalar relations. In particular, $\mathcal H_0=\mathbb C[z_a]/(f_\alpha)$, while higher groups record relations among these relations. A gauge projection is taken on the whole complex; for the reductive complexified groups used here, taking invariants is exact in characteristic zero.

Choose a positive grading for which $\deg\psi_\alpha=\deg f_\alpha$. At each degree the chain spaces are finite-dimensional, so their alternating dimension equals that of the cohomology. This explains why an index can be computed before solving for representatives, and also its limitation: it does not separate the individual $\mathcal H_j$. To construct secondaries we instead compute kernels, images and scalar multiplication on the surviving classes, and the Wess--Zumino examples in the main text and the restricted vector complex below use precisely this procedure.

\subsection{The Klebanov--Witten scalar quotient}
\label{app:kw}

The $SL(2)\times SL(2)$ action on a bifundamental is the vector action of $SO(4,\mathbb C)$, with its two-element kernel acting trivially. Using the epsilon tensors to convert the conjugate bifundamentals, the superpotential in Section~\ref{sec:kw} becomes the determinant of four vectors, and its derivatives impose rank at most two on their matrix. The invariant ring is
\begin{equation}
 R_{\rm KW}=\mathbb C[g_{ij}:1\leq i\leq j\leq4]/I_3(G),
\end{equation}
where $I_3(G)$ is generated by all three-by-three minors of the symmetric Gram matrix. The oriented four-vector invariant vanishes on the F-term variety. This ring has dimension seven and is Cohen--Macaulay~\cite{Kutz:1974}; this also follows from its description as a ring of $O(2,\mathbb C)$ invariants (Section~\ref{sec:kw}) and the existence result for invariant rings quoted in Section~\ref{sec:hironaka}~\cite{Hochster:1974}.

To exhibit a parameter choice, put $x=g_{33}$, $y=g_{34}$, $z=g_{44}$ and take the seven degree-two primaries
\begin{equation}\label{eq:kleb_prim}
\begin{aligned}
 p_1&=g_{11}+y-z,&p_2&=g_{12}+x-z,\\
 p_3&=g_{13}-x-y-z,&p_4&=g_{14}+x-y,\\
 p_5&=g_{22}+y-z,&p_6&=g_{23}+y,\\
 p_7&=g_{24}+x+z.&&
\end{aligned}
\end{equation}
In the primary quotient the matrix becomes
\begin{equation}
 G=\begin{pmatrix}
 -y+z&-x+z&x+y+z&-x+y\\
 -x+z&-y+z&-y&-x-z\\
 x+y+z&-y&x&y\\
 -x+y&-x-z&y&z
 \end{pmatrix}.
\end{equation}
Its ten distinct symmetric three-by-three minors span all ten cubic monomials in $x,y,z$: an exact coefficient calculation gives a nonzero determinant of magnitude $809$. Hence the quotient ideal is $(x,y,z)^3$, and a basis is
\begin{equation}
 1,\quad x,y,z,\quad x^2,xy,xz,y^2,yz,z^2.
\end{equation}
Finiteness of this quotient and the Cohen--Macaulay property establish freeness over the chosen seven primaries, as in the four-puncture example. Since $x,y,z$ have field degree two, the secondary profile is $1+3t^2+6t^4$, reproducing \eqref{eq:kw-series}.

\subsection{The restricted fermionic cohomology}
\label{app:restricted}

We now compute the invariant cohomology of \eqref{eq:restricted-Q}. Set $g_{ij}=v_i\cdot v_j$ for $i,j=1,2,3$, and let $S=\mathbb C[g_{11},g_{12},g_{13},g_{22},g_{23},g_{33}]$. The six Gram entries are algebraically independent before applying $Q$. Write $G=(g_{ij})$ and $\Delta=\det G$. Define
\begin{equation}
 u_i=v_i\cdot\psi,\qquad w=T\cdot\psi,\qquad
 U_1=u_2u_3,\quad U_2=u_3u_1,\quad U_3=u_1u_2.
\end{equation}
For exterior forms in the four components of $\psi$, use the same bilinear metric and orientation to define $*$. Put $B_i=*U_i$, $C_i=*u_i$, and $V=\psi_1\psi_2\psi_3\psi_4$. Invariant contractions give the following free $S$-bases of the chain spaces:
\begin{equation}
\begin{array}{c|l}
 j&\text{basis of }C_j^{SO(4)}\\\hline
 0&1\\
 1&u_1,u_2,u_3,w\\
 2&U_1,U_2,U_3,B_1,B_2,B_3\\
 3&C_1,C_2,C_3,*w\\
 4&V
\end{array}.
\end{equation}
The invariant tensors of $SO(4)$ are generated by the metric and the volume form, and their contractions yield precisely these elements after scalar Gram factors are removed. On the open set $\Delta\ne0$, the three $v_i$ together with $T$ form a basis of the vector space, which also shows independence over $S$.

With consistent exterior ordering, the nonzero differential blocks are
\begin{equation}
\label{eq:restricted-blocks}
 Qw=\Delta,\qquad
 QB_i=-\sum_j(\operatorname{adj}G)_{ji}u_j,\qquad
 QC_i=\sum_jG_{ji}U_j,\qquad QV=*w.
\end{equation}
All remaining basis elements are closed. Here $\operatorname{adj}G$ is the cofactor matrix, satisfying $G\operatorname{adj}G=\Delta\,\mathbf1$. Both $G$ and its adjugate are injective as matrices over the polynomial ring $S$, since their determinants are nonzero polynomials. The chain complex thus gives
\begin{equation}
 \mathcal H_0=S/(\Delta),\qquad
 \mathcal H_1=\operatorname{coker}(\operatorname{adj}G),\qquad
 \mathcal H_2=\operatorname{coker}G,\qquad
 \mathcal H_{j\geq3}=0.
\end{equation}
The cokernel means that the target generators are identified modulo the columns of the indicated matrix. The generators $u_i$ have degree four and the $U_i$ degree eight. Therefore
\begin{equation}
\begin{aligned}
 H_0(t)&=\frac{1-t^6}{(1-t^2)^6},&
 H_1(t)&=\frac{3t^4-3t^8}{(1-t^2)^6},&
 H_2(t)&=\frac{3t^8-3t^{10}}{(1-t^2)^6}
\end{aligned},
\end{equation}
which reduce to \eqref{eq:restricted-series}.

It remains to show that a single primary choice works in every fermion sector. Write
\begin{equation}
 G=\begin{pmatrix}c&d&a\\d&e&b\\a&b&f\end{pmatrix},\qquad
 p=f-d,\qquad P=\mathbb C[a,b,c,e,p].
\end{equation}
All five proposed primaries have degree two, and $S=P[d]$. As a polynomial in $d$, $\Delta$ has degree three with leading coefficient $-1$. The matrix $G$ has degree one with invertible leading coefficient
\begin{equation}
 L=\begin{pmatrix}0&1&0\\1&0&0\\0&0&1\end{pmatrix},
 \qquad \det L=-1,
\end{equation}
and $\operatorname{adj}G$ has degree two with leading coefficient $\operatorname{adj}L$. Polynomial division in $d$ therefore gives unique remainders, of degrees less than three, two and one, respectively. The common-primary bases are
\begin{equation}\label{eq:abjm_fer}
\begin{aligned}
 \mathcal H_0:&\quad 1,d,d^2,\\
 \mathcal H_1:&\quad u_1,u_2,u_3,du_1,du_2,du_3,\\
 \mathcal H_2:&\quad U_1,U_2,U_3.
\end{aligned}
\end{equation}
This proves that the entire invariant cohomology of the defined complex has a Hironaka decomposition over the five primaries with twelve secondaries.

\bibliographystyle{JHEP-notes}
\bibliography{hironaka_references}

@article{Kinney:2005ej,
    author = "Kinney, Justin and Maldacena, Juan Martin and Minwalla, Shiraz and Raju, Suvrat",
    title = "{An Index for 4 dimensional super conformal theories}",
    eprint = "hep-th/0510251",
    archivePrefix = "arXiv",
    doi = "10.1007/s00220-007-0258-7",
    journal = "Commun. Math. Phys.",
    volume = "275",
    pages = "209--254",
    year = "2007"
}

@article{Corley:2001zk,
    author = "Corley, Steve and Jevicki, Antal and Ramgoolam, Sanjaye",
    title = "{Exact correlators of giant gravitons from dual N=4 SYM theory}",
    eprint = "hep-th/0111222",
    archivePrefix = "arXiv",
    doi = "10.4310/ATMP.2001.v5.n4.a6",
    journal = "Adv. Theor. Math. Phys.",
    volume = "5",
    pages = "809--839",
    year = "2002"
}

@article{Berenstein:2004kk,
    author = "Berenstein, David",
    title = "{A Toy model for the AdS / CFT correspondence}",
    eprint = "hep-th/0403110",
    archivePrefix = "arXiv",
    doi = "10.1088/1126-6708/2004/07/018",
    journal = "JHEP",
    volume = "07",
    pages = "018",
    year = "2004"
}

@article{Chang:2013fba,
    author = "Chang, Chi-Ming and Yin, Xi",
    title = "{1/16 BPS states in $\mathcal N=$ 4 super-Yang-Mills theory}",
    eprint = "1305.6314",
    archivePrefix = "arXiv",
    primaryClass = "hep-th",
    doi = "10.1103/PhysRevD.88.106005",
    journal = "Phys. Rev. D",
    volume = "88",
    number = "10",
    pages = "106005",
    year = "2013"
}

@article{deMelloKoch:2025ngs,
    author = "de Mello Koch, Robert and Jevicki, Antal",
    title = "{Structure of loop space at finite N}",
    eprint = "2503.20097",
    archivePrefix = "arXiv",
    primaryClass = "hep-th",
    doi = "10.1007/JHEP06(2025)011",
    journal = "JHEP",
    volume = "06",
    pages = "011",
    year = "2025"
}

@article{deMelloKoch:2025rkw,
    author = "de Mello Koch, Robert and Jevicki, Antal",
    title = "{Hilbert space of finite N multi-matrix models}",
    eprint = "2508.11986",
    archivePrefix = "arXiv",
    primaryClass = "hep-th",
    doi = "10.1007/JHEP11(2025)145",
    journal = "JHEP",
    volume = "11",
    pages = "145",
    year = "2025"
}

@article{deMelloKoch:2025eqt,
    author = "de Mello Koch, Robert and Jevicki, Antal and Kemp, Garreth and Rudra, Anik",
    title = "{Collective theory at finite-N: reduction of the emergent Hilbert space}",
    eprint = "2510.22525",
    archivePrefix = "arXiv",
    primaryClass = "hep-th",
    doi = "10.1007/JHEP04(2026)048",
    journal = "JHEP",
    volume = "04",
    pages = "048",
    year = "2026"
}

@article{deMelloKoch:2026secondary,
    author = {de Mello Koch, Robert and Rodrigues, Jo{\~a}o P.},
    title = {Secondary invariants and non-perturbative states},
    eprint = {2604.15600},
    archivePrefix = {arXiv},
    primaryClass = {hep-th},
    doi = {10.1007/JHEP06(2026)251},
    journal = {JHEP},
    volume = {06},
    pages = {251},
    year = {2026}
}

@article{Cachazo:2002ry,
    author = "Cachazo, Freddy and Douglas, Michael R. and Seiberg, Nathan and Witten, Edward",
    title = "{Chiral rings and anomalies in supersymmetric gauge theory}",
    eprint = "hep-th/0211170",
    archivePrefix = "arXiv",
    doi = "10.1088/1126-6708/2002/12/071",
    journal = "JHEP",
    volume = "12",
    pages = "071",
    year = "2002"
}

@article{Aharony:2008ug,
    author = "Aharony, Ofer and Bergman, Oren and Jafferis, Daniel Louis and Maldacena, Juan",
    title = "{N=6 superconformal Chern-Simons-matter theories, M2-branes and their gravity duals}",
    eprint = "0806.1218",
    archivePrefix = "arXiv",
    primaryClass = "hep-th",
    doi = "10.1088/1126-6708/2008/10/091",
    journal = "JHEP",
    volume = "10",
    pages = "091",
    year = "2008"
}

@article{Cremonesi:2016nbo,
    author = "Cremonesi, Stefano and Mekareeya, Noppadol and Zaffaroni, Alberto",
    title = "{The moduli spaces of 3d $ \mathcal{N}\ge 2 $ Chern-Simons gauge theories and their Hilbert series}",
    eprint = "1607.05728",
    archivePrefix = "arXiv",
    primaryClass = "hep-th",
    doi = "10.1007/JHEP10(2016)046",
    journal = "JHEP",
    volume = "10",
    pages = "046",
    year = "2016"
}

@article{Kim:2009wb,
    author = "Kim, Seok",
    title = "{The Complete superconformal index for N=6 Chern-Simons theory}",
    eprint = "0903.4172",
    archivePrefix = "arXiv",
    primaryClass = "hep-th",
    doi = "10.1016/j.nuclphysb.2009.06.025",
    journal = "Nucl. Phys. B",
    volume = "821",
    pages = "241--284",
    year = "2009",
    note = "[Erratum: Nucl.Phys.B 864, 884 (2012)]"
}

@article{Choi:2023znd,
    author = "Choi, Sunjin and Kim, Seok and Lee, Eunwoo and Lee, Siyul and Park, Jaemo",
    title = "{Towards quantum black hole microstates}",
    eprint = "2304.10155",
    archivePrefix = "arXiv",
    primaryClass = "hep-th",
    doi = "10.1007/JHEP11(2023)175",
    journal = "JHEP",
    volume = "11",
    pages = "175",
    year = "2023",
    note = "[Erratum: JHEP 03, 091 (2025)]"
}

@article{Choi:2023vdm,
    author = "Choi, Jaehyeok and Choi, Sunjin and Kim, Seok and Lee, Jehyun and Lee, Siyul",
    title = "{Finite N black hole cohomologies}",
    eprint = "2312.16443",
    archivePrefix = "arXiv",
    primaryClass = "hep-th",
    doi = "10.1007/JHEP12(2024)029",
    journal = "JHEP",
    volume = "12",
    pages = "029",
    year = "2024"
}

@article{Chang:2024zqi,
    author = "Chang, Chi-Ming and Lin, Ying-Hsuan",
    title = "{Holographic covering and the fortuity of black holes}",
    eprint = "2402.10129",
    archivePrefix = "arXiv",
    primaryClass = "hep-th",
    month = "2",
    year = "2024"
}

@article{Gadde:2025yoa,
    author = "Gadde, Abhijit and Lee, Eunwoo and Raj, Rajat and Tomar, Shivansh",
    title = "{Probing non-graviton spectra in $ \mathcal{N}=4 $ SYM via BMN truncation and S-duality}",
    eprint = "2506.13887",
    archivePrefix = "arXiv",
    primaryClass = "hep-th",
    doi = "10.1007/JHEP02(2026)026",
    journal = "JHEP",
    volume = "02",
    pages = "026",
    year = "2026"
}

@article{Chang:2022mjp,
    author = "Chang, Chi-Ming and Lin, Ying-Hsuan",
    title = "{Words to describe a black hole}",
    eprint = "2209.06728",
    archivePrefix = "arXiv",
    primaryClass = "hep-th",
    doi = "10.1007/JHEP02(2023)109",
    journal = "JHEP",
    volume = "02",
    pages = "109",
    year = "2023"
}

@article{Chang:2025mqp,
    author = "Chang, Chi-Ming and Lin, Ying-Hsuan",
    title = "{Violation of S-duality in classical $Q$-cohomology}",
    eprint = "2510.24008",
    archivePrefix = "arXiv",
    primaryClass = "hep-th",
    month = "10",
    year = "2025"
}

@article{Grant:2008sk,
    author = "Grant, Lars and Grassi, Pietro A. and Kim, Seok and Minwalla, Shiraz",
    title = "{Comments on 1/16 BPS Quantum States and Classical Configurations}",
    eprint = "0803.4183",
    archivePrefix = "arXiv",
    primaryClass = "hep-th",
    doi = "10.1088/1126-6708/2008/05/049",
    journal = "JHEP",
    volume = "05",
    pages = "049",
    year = "2008"
}

@article{Choi:2022caq,
    author = "Choi, Sunjin and Kim, Seok and Lee, Eunwoo and Park, Jaemo",
    title = "{The shape of non-graviton operators for SU(2)}",
    eprint = "2209.12696",
    archivePrefix = "arXiv",
    primaryClass = "hep-th",
    doi = "10.1007/JHEP09(2024)029",
    journal = "JHEP",
    volume = "09",
    pages = "029",
    year = "2024"
}

@article{Stanley:1979,
 author = {Stanley, Richard P.},
 title = {Invariants of finite groups and their applications to combinatorics},
 journal = {Bull. Amer. Math. Soc.},
 volume = {1},
 number = {3},
 pages = {475--511},
 year = {1979},
 doi = {10.1090/S0273-0979-1979-14597-X}
}

@article{Chevalley:1955,
 author = {Chevalley, Claude},
 title = {Invariants of finite groups generated by reflections},
 journal = {Amer. J. Math.},
 volume = {77},
 number = {4},
 pages = {778--782},
 year = {1955},
 doi = {10.2307/2372597}
}

@article{Gadde:2011uv,
    author = "Gadde, Abhijit and Rastelli, Leonardo and Razamat, Shlomo S. and Yan, Wenbin",
    title = "{Gauge Theories and Macdonald Polynomials}",
    eprint = "1110.3740",
    archivePrefix = "arXiv",
    primaryClass = "hep-th",
    reportNumber = "YITP-SB-11-30",
    doi = "10.1007/s00220-012-1607-8",
    journal = "Commun. Math. Phys.",
    volume = "319",
    pages = "147--193",
    year = "2013"
}

@article{Beauville_2000,
   title={Symplectic singularities},
   volume={139},
   ISSN={1432-1297},
   url={http://dx.doi.org/10.1007/s002229900043},
   DOI={10.1007/s002229900043},
   number={3},
   journal={Inventiones Mathematicae},
   publisher={Springer Science and Business Media LLC},
   author={Beauville, Arnaud},
   year={2000},
   month=mar, pages={541--549} }

@article{Minahan:1996fg,
    author = "Minahan, Joseph A. and Nemeschansky, Dennis",
    title = "{An N=2 superconformal fixed point with E(6) global symmetry}",
    eprint = "hep-th/9608047",
    archivePrefix = "arXiv",
    reportNumber = "USC-96-18",
    doi = "10.1016/S0550-3213(96)00552-4",
    journal = "Nucl. Phys. B",
    volume = "482",
    pages = "142--152",
    year = "1996"
}

@article{Benvenuti_2010,
 author = {Benvenuti, Sergio and Hanany, Amihay and Mekareeya, Noppadol},
 title = {The {Hilbert} series of the one instanton moduli space},
 journal = {JHEP}, volume = {06}, pages = {100}, year = {2010},
 doi = {10.1007/JHEP06(2010)100},
 eprint = {1005.3026}, archivePrefix = {arXiv}, primaryClass = {hep-th}
}

@article{Hanany_2011,
 author = {Hanany, Amihay and Mekareeya, Noppadol},
 title = {Tri-vertices and {SU(2)}'s},
 journal = {JHEP}, volume = {02}, pages = {069}, year = {2011},
 doi = {10.1007/JHEP02(2011)069},
 eprint = {1012.2119}, archivePrefix = {arXiv}, primaryClass = {hep-th}
}

@article{Klebanov_1998,
 author = {Klebanov, Igor R. and Witten, Edward},
 title = {Superconformal field theory on threebranes at a {Calabi--Yau} singularity},
 journal = {Nucl. Phys. B}, volume = {536}, pages = {199--218}, year = {1998},
 doi = {10.1016/S0550-3213(98)00654-3},
 eprint = {hep-th/9807080}, archivePrefix = {arXiv}, primaryClass = {hep-th}
}

@article{Hochster:1974,
 author = {Hochster, Melvin and Roberts, Joel L.},
 title = {Rings of invariants of reductive groups acting on regular rings are {Cohen--Macaulay}},
 journal = {Adv. Math.}, volume = {13}, pages = {115--175}, year = {1974}
}

@article{Hochster:1972,
 author = {Hochster, Melvin},
 title = {Rings of invariants of tori, {Cohen--Macaulay} rings generated by monomials, and polytopes},
 journal = {Ann. of Math.}, volume = {96}, pages = {318--337}, year = {1972}
}

@article{Stanley:1978,
 author = {Stanley, Richard P.},
 title = {Hilbert functions of graded algebras},
 journal = {Adv. Math.}, volume = {28}, pages = {57--83}, year = {1978}
}

@article{Watanabe:1974a,
 author = {Watanabe, Keiichi},
 title = {Certain invariant subrings are {Gorenstein}. {I}},
 journal = {Osaka J. Math.}, volume = {11}, pages = {1--8}, year = {1974}
}

@article{Watanabe:1974b,
 author = {Watanabe, Keiichi},
 title = {Certain invariant subrings are {Gorenstein}. {II}},
 journal = {Osaka J. Math.}, volume = {11}, pages = {379--388}, year = {1974}
}

@article{Kutz:1974,
 author = {Kutz, Ronald E.},
 title = {{Cohen--Macaulay} rings and ideal theory in rings of invariants of algebraic groups},
 journal = {Trans. Amer. Math. Soc.}, volume = {194}, pages = {115--129}, year = {1974}
}

@article{Budzik:2025zvu,
    author = "Budzik, Kasia and Kulp, Justin",
    title = "{Loop corrected supercharges from holomorphic anomalies}",
    eprint = "2512.07771",
    archivePrefix = "arXiv",
    primaryClass = "hep-th",
    doi = "10.1007/JHEP08(2026)089",
    journal = "JHEP",
    volume = "08",
    pages = "089",
    year = "2026"
}
\end{document}